\documentclass[12pt]{article}

\usepackage[utf8]{inputenc}
\usepackage[T1]{fontenc}
\usepackage{lmodern}
\usepackage{geometry}
\usepackage{graphicx}
\usepackage{longtable}
\usepackage{booktabs}
\usepackage{array}
\usepackage{calc}
\usepackage{textcomp}
\usepackage{amsmath}
\usepackage{amssymb}
\usepackage{etoolbox}
\usepackage{natbib}
\usepackage{hyperref}
\usepackage{tikz}
\usetikzlibrary{arrows.meta, positioning, calc, shapes.geometric, fit}
\hypersetup{hidelinks}

\newcommand{\ind}{\mathbf{1}}

\makeatletter
\patchcmd\LT@output{\copy\LT@foot\vss}{\copy\LT@foot\vfil}{}{}
\patchcmd\LT@output{\copy\LT@foot\vss}{\copy\LT@foot\vfil}{}{}
\makeatother

\title{Temporal Dynamics of Development Aid in Africa: Evidence from a Staggered Difference-in-Differences Study of China and World Bank Projects}

\title{
Temporal Dynamics of Development Aid in Africa: Evidence from a Staggered Difference-in-Differences Study of China and World Bank Projects
}
\author{Mattias Antar\thanks{ PhD Student, Division of Ageing and Social Change (ASC), Department of Culture and Society (IKOS), Linköping University, SE-581 83 Linköping, Sweden. Email: \texttt{mattias.antar@liu.se}. Division of Ageing and Social Change, Linköping University, SE-581 83 Linköping, Sweden. Phone: +46 11 36 35 35.}
\and Adel Daoud\thanks{Professor of Computational Social Science, Institute for Analytical Sociology, Linköping University, 601 74 Norrköping, Sweden. Email: \texttt{adel.daoud@liu.se}.}
\and Connor Jerzak\thanks{Assistant Professor, Department of Government, University of Texas at Austin, 158 W 21st St Stop A1800, Austin, TX 78712, USA. Email: \texttt{connor.jerzak@austin.utexas.edu}.}}
\date{}

\begin{document}

\maketitle

\begin{abstract}
\noindent Subnational studies of aid effectiveness often rely on repeated cross-sections or nighttime lights, making it difficult to separate local treatment effects from baseline differences and potentially favoring infrastructure-heavy projects. We address these limitations by studying World Bank and Chinese development projects in Africa with a balanced panel of 2,166 DHS clusters across 35 countries from 2002 to 2013. Geocoded AidData projects are linked to satellite-imputed International Wealth Index estimates, a household-centered measure of material living standards. We compare a conventional two-way fixed effects (TWFE) event-study with the switcher--stayer estimator of de Chaisemartin and D'Haultf\oe{}uille (dCdH), which avoids contaminated comparisons under staggered treatment timing. Pre-treatment diagnostics show that project placement is frequently selective: clusters that later receive projects often begin from weaker relative positions before treatment onset. Consequently, TWFE often implies larger post-treatment gains than the preferred staggered-treatment design supports. Under dCdH, the evidence becomes more selective and sector-specific. For the World Bank, positive evidence is strongest in Health, while Education shows positive but less cleanly identified gains. For China, Water Supply and Sanitation and Other Social Infrastructure and Services show positive associations with local wealth, although residual selection concerns remain. By contrast, Chinese Energy Generation and Supply appears strongly positive under TWFE but falls close to zero under dCdH. Overall, the results do not support a donor-wide claim that either the World Bank or China uniformly improves local wealth. Instead, estimated effects are concentrated in a limited set of donor--sector panels and depend strongly on how treatment timing, selection, and outcome measurement are handled.


\end{abstract}

\clearpage
\section{Introduction}\label{introduction}

Whether foreign aid improves economic conditions remains a persistent and disputed question in development economics. The local level is especially important because projects are implemented there, households experience gains or disruptions, and donor portfolios may translate into welfare through very different channels. A large macro-level literature has studied the relationship between aid and national growth, yet the empirical record remains inconclusive because aid is endogenous, recipient countries are highly heterogeneous, and aggregate outcomes may obscure localized impacts \citep{BurnsideDollar2000Aid, EasterlyLevineRoodman2004Aid, RajanSubramanian2008Aid, Dreher2021AidChinaGrowth}. 

In response, the literature has increasingly shifted from country-level aggregates to subnational analyses that link geocoded projects to local outcomes, with the aim of observing whether aid changes economic conditions where it is actually delivered \citep{Bluhm2018Connective, Dreher2021AidChinaGrowth, Gehring2022ChinaWB}. This ``subnational turn'' has produced more localized evidence than earlier cross-country regressions, but it also leaves a specific gap: much subnational work is cross-sectional or repeated cross-sectional, making it difficult to distinguish treatment effects from baseline differences across places, and much of it uses nighttime lights, an outcome proxy that may mechanically favor infrastructure-heavy donors and sectors. This paper addresses both problems simultaneously by combining a neighborhood-level wealth panel with a staggered difference-in-differences design that places selection bias and heterogeneous treatment timing at the center of the empirical design.

The outcome data problem is particularly relevant in sub-Saharan Africa. Household surveys remain the gold standard for measuring living conditions, but they are expensive, geographically incomplete, and not designed to repeatedly observe the same local units over time. For that reason, many subnational studies have relied on nighttime lights as a proxy for economic activity. Nighttime lights have clear advantages, but it is an imperfect measure of household welfare, especially in low-electrification or low-density settings where improvements in housing quality, assets, sanitation, or water access may occur without a corresponding increase in luminosity \citep{Jean2016SatelliteML, Bluhm2018Connective, Pettersson2023TimeseriesImagery}. In aid settings, that limitation is consequential: a proxy that is especially sensitive to electrification and large infrastructure may mechanically favor some sectors over others and may understate welfare gains from interventions whose effects operate through health, education, or household asset accumulation.

This paper addresses that outcome data problem by using the satellite-derived International Wealth Index (IWI) surface developed by \citet{Pettersson2023TimeseriesImagery}. The IWI is an asset-based measure of household living standards, rooted in ownership of durable goods and housing and service characteristics, and therefore sits closer to multidimensional material well-being than luminosity-based proxies \citep{SmitsSteendijk2015IWI, Pettersson2023TimeseriesImagery}. We link geocoded aid projects from AidData to a balanced panel of 2,166 Demographic and Health Survey (DHS) clusters across 35 African countries, observed over four three-year periods from 2002--2004 through 2011--2013. Throughout the paper, we treat DHS clusters as neighborhood-level units: substantively, they approximate villages in rural areas and neighborhoods in urban settings, while empirically they provide the stable local reference points needed to study within-location change over time.

The paper focuses on two major external development financiers in Africa during the study period: the World Bank and China. This comparison is informative because the two donors embody distinct portfolios, implementation styles, and theories of change. The World Bank's comparative advantage is often argued to lie in social sectors, institution-building, and human-capital investments; China's is usually associated with transport, energy, utilities, and other visible infrastructure \citep{Andersen2006USPolitics, Strange2017Tracking, Dreher2022Banking, Gehring2022ChinaWB}. 

Yet the existing comparative literature still leaves an important gap. Much comparative evidence remains at the country level, and the subnational work that directly compares China and the World Bank has typically used ADM-level region-year fixed-effects or IV designs and conflict/stability outcomes rather than balanced neighborhood-level welfare panels with staggered treatment timing, leaving open how donor--sector conclusions change when identification is driven by within-neighborhood changes over time \citep{Dreher2021AidChinaGrowth, Gehring2022ChinaWB}. A closely related recent study uses the same underlying wealth surface in a continent-scale, sector-specific design and strengthens identification with image-augmented inverse-probability weighting \citep{daoud2026chinese}. That design is useful for modeling selection on observed and image-derived confounders, but it does not make within-neighborhood treatment timing the primary source of identifying variation. Our contribution is therefore complementary: we use the same welfare outcome to study staggered adoption within a balanced panel, asking how donor--sector conclusions change when identification is driven by changes within the same neighborhoods over time and when conventional TWFE estimates are compared with a switcher--stayer dCdH design. Put differently, the distinction is not the outcome data alone but the way temporal variation is used: Daoud et al. emphasize image-augmented adjustment for assignment, whereas this paper foregrounds timing, switchers, and within-neighborhood event-study dynamics.

That identification challenge is important. Development projects are not randomly scattered across space. Donors respond to need, visibility, access, political pressure, and recipient-side bargaining, among other factors \citep{OhlerNunnenkamp2014Needs, Dreher2019Leaders, Jablonski2014AidVotes}. In this setting, the standard two-way fixed effects (TWFE) event-study estimator is vulnerable for two distinct reasons. 

First, the contrast between the last untreated period and the omitted onset period can reveal patterns consistent with selection bias; in our four-period setting, that diagnostic is the pre-treatment diagnostic contrast at $\ell=-1$. In this setting, the standard two-way fixed effects (TWFE) event-study estimator is vulnerable for two distinct reasons. First, the contrast between the last untreated period and the omitted onset period can reveal patterns consistent with selection bias; in our four-period setting, that diagnostic is the pre-treatment diagnostic contrast at $\ell=-1$. We define the \textit{pre-treatment contrast} as the difference in outcome trajectories between switching clusters and untreated comparison clusters in the period immediately preceding treatment onset ($\ell=-1$). A non-zero contrast indicates selective project placement---meaning treated areas were already experiencing differential wealth dynamics prior to the intervention---which threatens the parallel trends assumption and can severely bias conventional difference-in-differences estimates. 

Second, with staggered treatment timing and heterogeneous effects, TWFE combines many implicit two-by-two comparisons, some of which receive negative weights and can therefore distort dynamic treatment profiles \citep{deChaisemartin2020TWFE, Callaway2021DiD}. Both concerns prove empirically consequential in this application, as the diagnostic evidence in Section~\ref{pretrends} demonstrates.

We therefore use TWFE as a diagnostic benchmark and the switcher--stayer estimator of \citet{deChaisemartin2020TWFE} as our preferred dCdH specification. In our analysis, we estimate the model separately across 23 donor--sector panels. This disaggregation is important substantively, because different sectors plausibly operate through different channels and over different horizons, and methodologically, because it avoids masking heterogeneity behind a single average effect. The central comparison is between conventional staggered-TWFE conclusions and what remains under the preferred switcher--stayer dCdH specification, a distinction that cuts across both donors.

Our results show that estimator choice has substantive consequences in this setting. Across donor--sector panels, pre-treatment diagnostic contrasts frequently show patterns consistent with selection bias, with treated neighborhoods often starting from weaker relative positions before onset. In those settings, TWFE tends to report larger post-treatment gains. Once we move to the preferred switcher--stayer dCdH estimator, the substantive picture becomes more selective and more conditional on identification quality. The evidence therefore supports a narrow directional conclusion: estimated local aid effects vary across panels, donors, and estimators, and the strongest evidence is concentrated in a limited subset of cases with comparatively small preferred dCdH pre-treatment diagnostic contrasts.

The paper makes three contributions. First, it brings a local wealth panel into the aid-effectiveness literature by combining geocoded projects with a satellite-imputed, temporally consistent measure of household material well-being. Second, it demonstrates that selection and staggered-treatment biases are central challenges in subnational aid evaluation. Third, it provides a donor--sector comparison centered on which types of interventions are associated with within-neighborhood wealth gains over the observed post-treatment window.

The remainder of the paper proceeds as follows. Section~\ref{background} summarizes the institutional differences between the World Bank and China, explains why selection bias in treatment timing is plausible, and motivates the use of a satellite-derived wealth outcome. The next section describes the data and empirical design. We then present diagnostic evidence from pre-treatment diagnostic contrasts, compare TWFE and dCdH estimates, and discuss robustness materials before concluding.

\section{Institutional background}\label{background}

\subsection{Donor logics and expected timing of effects}

The World Bank and China finance development through different institutional logics, and those differences generate distinct empirical expectations at the local level. The World Bank's development model is rooted in a broad poverty-reduction mandate and is often implemented through social-sector and public-service interventions, typically alongside governance, administrative, or institutional objectives \citep{Andersen2006USPolitics, Hernandez2017NewDonors}. Even when the Bank finances infrastructure, its portfolio is comparatively more embedded in public systems and more likely to be tied to service delivery, regulation, or implementation through recipient institutions. This implies that the local effects of World Bank projects may be slower to materialize in household wealth data than the effects of highly visible physical infrastructure. Health, education, and water-related projects, for example, may require time for facilities to become operational, staff to be deployed, and household behavior to adjust before gains appear in asset-based welfare measures.

The economic channel is also likely to be indirect. Health projects can raise IWI if improved services reduce out-of-pocket medical spending, protect adult labor supply, or improve child health in ways that allow households to accumulate durables or improve housing and sanitation over time. Education projects may operate through school attendance, parental labor allocation, and expectations about returns to human capital, so their effects on an asset-based wealth index should also emerge gradually rather than immediately.

China's model differs in both composition and execution. Chinese development finance in Africa has emphasized transport, energy, utilities, and other forms of ``hard'' infrastructure, often under a state-led and non-interference framework \citep{Strange2017Tracking, Dreher2022Banking}. The boundary between concessional aid, commercial finance, and strategic overseas investment is also blurrier in the Chinese case than in traditional multilateral lending, which makes the on-the-ground portfolio more heterogeneous \citep{Strange2017Tracking, Malik2021BeltRoad}. In principle, that model can generate faster local effects when projects relax binding constraints on mobility, electricity, or access to water. 

At the same time, the household-level consequences need not be uniformly positive. Energy projects may raise firm productivity, lighting, and commercial activity, which nighttime lights capture relatively well, without quickly changing household asset ownership if the gains accrue mainly to firms, public facilities, or grid-connected corridors. Transport projects can lower trade costs and improve market access, but road corridors may also bypass nearby communities, shift activity toward larger nodes, or disrupt local labor and retail markets during and after construction. Large infrastructure can therefore raise measured economic activity without immediately improving household assets, and some projects may primarily benefit transit corridors, firms, or politically connected intermediaries rather than nearby households.

These institutional differences shape the empirical expectations for timing and sign of event-study estimates. If the World Bank's comparative advantage lies in human capital and service delivery, we should expect any improvements in household wealth to emerge gradually, especially in Health and Education. If China's comparative advantage lies in utilities and connective infrastructure, some gains may appear more quickly where projects directly improve service access, while other sectors may show weaker household-level responses within the time horizon. These are empirical expectations, motivating a sector-by-sector design rather than a pooled donor comparison.

\subsection{Selection bias}

A second institutional feature of the aid environment is that project placement is inherently selective. Donors do not allocate projects to an arbitrary spatial sample of communities. Within recipient countries, project locations are shaped by a mixture of poverty targeting, administrative feasibility, political bargaining, and donor strategy \citep{OhlerNunnenkamp2014Needs, Dreher2019Leaders}. For the World Bank, even explicitly pro-poor mandates do not imply uniform placement in the poorest localities. Project implementation often favors places with stronger administrative capacity, lower logistical cost, or better connectivity, and some work proposes that World Bank portfolios do not always map cleanly onto the bottom of the local welfare distribution \citep{OhlerEtAl2019Bottom40, Briggs2012Electrifying, Yanguas2015Barriers}. In fragile environments, implementation may also be delegated away from central state structures, further complicating the relationship between need and observed placement \citep{Chasukwa2019Institutional, Marchesi2021Delegation}.

Selection bias is at least as plausible for China, though the mechanism may differ. A growing literature shows that Chinese projects respond to political and strategic incentives, including leader favoritism, commercial opportunity, and executive bargaining \citep{Dreher2019Leaders, Strange2017Tracking}. In that sense, Chinese placement may be less tightly linked to conventional poverty metrics and more strongly tied to state-level priorities, bilateral relationships, or politically salient geographies. The same location may also receive multiple projects across sectors or financiers, creating a form of clustering that complicates attribution if not explicitly acknowledged.

For empirical work, the implication is straightforward: treated and untreated neighborhoods are unlikely to be comparable without strong design assumptions. Moreover, selection bias need not work in only one direction. Some projects may be directed toward distressed areas and exhibit negative pre-treatment diagnostic contrasts; others may favor already advantaged, accessible, or politically central locations and exhibit positive values of the pre-treatment diagnostic contrast. Either pattern can bias conventional difference-in-differences estimates. In our setting, selection bias is one of the substantive reasons that a benchmark TWFE estimate and a preferred staggered-treatment estimator can tell materially different stories about the same donor--sector panel.

\subsection{The International Wealth Index as an outcome}

The third piece of background concerns outcome measurement. Much of the subnational aid literature has used nighttime lights because it is widely available, comparable across space, and plausibly responsive to local economic activity. For many questions, that choice is sensible. But in the context of aid effectiveness, nighttime lights are better understood as a proxy for electrification, built-up intensity, and certain forms of commercial activity than as a comprehensive measure of household welfare \citep{Jean2016SatelliteML, Bluhm2018Connective}. This distinction is especially important in Africa, where low baseline electrification and sparse settlement can make welfare changes difficult to detect in luminosity data.

The International Wealth Index is closer to the welfare concept we care about in this paper. Developed as a standardized cross-country measure of material living standards, the IWI aggregates information on ownership of consumer durables, housing materials, sanitation, water access, electricity, and crowding \citep{SmitsSteendijk2015IWI}. \citet{Pettersson2023TimeseriesImagery} use DHS-based ground truth from 138 surveys and 57,195 survey clusters---covering roughly 1.2 million households---to train a temporally aware deep-learning model that predicts IWI from multi-temporal satellite imagery. The resulting data product provides a harmonized wealth surface over Africa at approximately 6.72 $\times$ 6.72 km resolution in three-year intervals. Relative to repeated cross-sectional surveys alone, that greatly expands the set of locations and periods for which neighborhood-level wealth can be observed.

In this paper, the role of earth observation is therefore primarily to supply a pre-processed, spatially and temporally consistent IWI surface derived from satellite imagery and DHS training data for downstream causal analysis; we do not analyze raw satellite imagery directly. Remote sensing does not by itself resolve the identification problem; credible causal interpretation still depends on treatment assignment and the difference-in-differences design \citep{Jerzak2023EOandCausal, Pettersson2023TimeseriesImagery}. This distinction guides interpretation. A better outcome measure does not substitute for credible identification, but it can materially improve what we mean by local welfare change. In particular, it allows us to evaluate donor--sector interventions against a household-centered measure of living standards rather than a proxy that mechanically privileges visible infrastructure. For that reason, the IWI is especially well suited to a paper whose aim is to compare local wealth dynamics across donors and sectors rather than indexing short-run commercial activity.

\section{Data}\label{data}

This study combines geocoded development projects with satellite-imputed
measures of local material well-being to construct a balanced panel for
difference-in-differences analysis. The key design choice is to work at
the \emph{DHS cluster-period} level. Each observation corresponds to a
Demographic and Health Survey (DHS) cluster observed in a given
three-year period; substantively, a DHS cluster approximates a village
in rural areas and a neighborhood in urban areas
\citep{Pettersson2023TimeseriesImagery}. This unit of observation is
small enough to study localized project exposure while still being
comparable across countries and periods.

The empirical contribution of earth observation in this paper is on the \emph{outcome} side. We do not use raw satellite images to estimate treatment propensities or to proxy for map-based placement criteria, as in image-augmented assignment models. Instead, we use the satellite-imputed International Wealth Index (IWI) surface developed by \citet{Pettersson2023TimeseriesImagery} to obtain a spatially and temporally consistent measure of local living standards that can be linked to geocoded aid projects. In the terminology of recent EO--ML workflows, this paper is best understood as using \emph{outcome imputation for downstream causal analysis} rather than EO image deconfounding \citep{Jerzak2023EOandCausal}.

We do not implement image-augmented treatment-assignment modeling in this paper; that approach requires a distinct high-dimensional modeling layer that would complicate rather than complement the within-neighborhood event-study design pursued here. We therefore use the satellite model solely to obtain the outcome surface, reserving image-augmented assignment for complementary designs such as \citet{daoud2026chinese}.

\begin{figure}[htbp]
\centering
\includegraphics[width=0.78\linewidth]{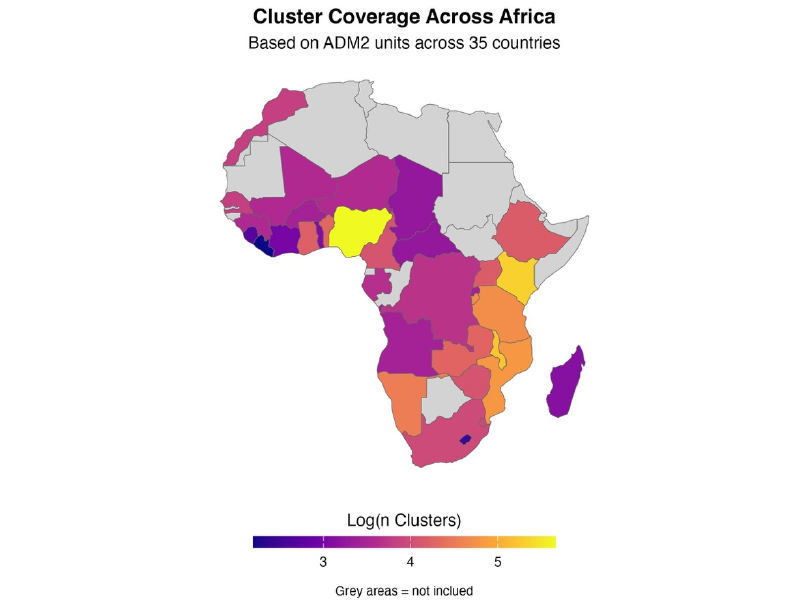}
\caption{Spatial coverage of the balanced sample across the 35-country
study area. Darker shading indicates a larger number of sampled
clusters. Sample density varies across countries, which is one
reason sparse donor--sector panels warrant cautious interpretation.
Country-level counts are reported in Appendix Table~\ref{tab:coverage}.
\textit{Note:} Map lines delineate study areas and do not necessarily
depict accepted national boundaries.}
\label{fig:coverage}
\end{figure}

The final analytic sample contains 2,166 DHS clusters across 35 African
countries, observed over four consecutive three-year periods:
2002--2004, 2005--2007, 2008--2010, and 2011--2013. This yields a
balanced panel of 8,664 cluster-period observations. Each cluster is
followed over time through a stable cluster identifier, which
allows us to estimate within-cluster changes in wealth rather than
relying on repeated cross-sectional differences alone. Figure~\ref{fig:coverage} shows
the spatial coverage of the sample across the 35-country study area, and
Appendix Table~\ref{tab:coverage} reports the country composition.

\subsection{Unit of observation and sample construction}\label{unit}

The unit of observation is the DHS \emph{cluster-period}. This distinction is important because several spatial objects enter the data construction. First, the DHS cluster is the underlying survey-based local unit. Second, the Pettersson et al.\ wealth surface is generated on 6.72 $\times$ 6.72 km image tiles and reported as three-year median surface values. Third, the aid data are recorded as project locations with varying geocoding precision. These objects are not treated as interchangeable. The observed panel unit is always the DHS cluster-period; the IWI outcome is assigned to that unit from the corresponding 6.72 km tile; and treatment is defined by proximity from the DHS cluster centroid to a reported project location.

This formulation is consistent with the underlying DHS-based wealth model. In \citet{Pettersson2023TimeseriesImagery}, the cluster is the primary sampling unit of the DHS and corresponds to a village or urban neighborhood. The EO--ML model predicts average cluster-level material wealth from multi-temporal satellite imagery and produces a harmonized wealth surface for Africa at approximately 6.72 km resolution in three-year median intervals from 1990 to 2019. In our application, the satellite-derived outcome enters in tabular form at the DHS-cluster level, allowing those same local units to be observed repeatedly across periods \citep{Pettersson2023TimeseriesImagery}.

The main analysis is restricted to 2002--2013 because the donor--sector panel data only identify treatment status during these four three-year median waves: 2002--2004, 2005--2007, 2008--2010, and 2011--2013. In other words, while the satellite-imputed IWI surface is available for a much longer period, from 1990 to 2019, the panel data used in this study only allow each DHS cluster-period to be classified as treated or untreated within the 2002--2013 window. The baseline difference-in-differences analysis is therefore conducted on this common four-period panel. Restricting the estimation sample in this way ensures that treatment status is defined consistently across all 23 donor--sector panels.

Because these wealth values are imputed rather than directly observed in repeated household surveys, measurement error is a substantive concern. We therefore return to this issue in the measurement-error sensitivity analysis,
which assesses whether the main donor--sector patterns are sensitive to plausible noise in the satellite-imputed IWI outcome.

\subsection{Outcome: Satellite-imputed local wealth}\label{outcome}

The outcome variable is the International Wealth Index (IWI), a continuous measure of material living standards bounded between 0 and 100 \citep{SmitsSteendijk2015IWI}. The IWI is constructed from household asset ownership and housing characteristics, including electricity, sanitation, water access, consumer durables, flooring, and sleeping arrangements. It is therefore closer to household material well-being than proxies based solely on luminosity or built-up intensity.

We use the temporally consistent African IWI surface developed by
\citet{Pettersson2023TimeseriesImagery}. Their EO--ML pipeline is
trained on approximately 1.2 million households in 57,195 DHS clusters
drawn from 138 nationally representative DHS surveys across 36 African
countries. The predictive task is to estimate the mean cluster-level IWI
from a sequence of satellite images centered on the survey cluster.
Their model combines a convolutional image encoder with a recurrent
temporal component, allowing it to learn both spatial and temporal
features associated with evolving local living conditions
\citep{Pettersson2023TimeseriesImagery}. The resulting product is a
harmonized wealth surface for Africa at 6.72 $\times$ 6.72 km
resolution in three-year intervals from 1990 to 2019.

In this paper, we assign to each DHS cluster-period the corresponding
satellite-imputed IWI value, denoted $Y_{it}$, where $i$ indexes the
cluster and $t$ indexes the three-year period. Operationally, the outcome
therefore enters as a cluster-level imputed welfare measure rather than
as raw imagery. The interpretation is therefore specific to the causal design. The
satellite model expands the spatial and temporal reach of the outcome
data, but it does not itself identify causal effects. Causal
interpretation still depends on the treatment definition and the
difference-in-differences design.

The main advantage of this outcome is that it converts otherwise sparse,
repeated cross-sectional survey information into a panel-like local
welfare series. This is especially valuable in a context where the same
neighborhoods are not repeatedly surveyed by the DHS. At the same time,
the outcome is model-generated and therefore subject to prediction
error. We therefore interpret IWI as a noisy but systematically
constructed measure of neighborhood-level material well-being. To the
extent that some component of the prediction error is stable within
cluster or common across periods, fixed effects will absorb part of that
variation; to the extent that it remains idiosyncratic, it will reduce
precision and attenuate estimated treatment effects
\citep{Pettersson2023TimeseriesImagery}.

\subsection{Treatment: Geocoded aid projects}\label{treatment}

Treatment data come from two AidData releases. World Bank projects are
drawn from the World Bank Geocoded Research Release Level 1 v1.4.2
\citep{AidData2017WBGeocoded}. Chinese projects are drawn from AidData's
Global Chinese Development Finance Dataset v1.1.1, assembled using the
Tracking Underreported Financial Flows (TUFF) methodology
\citep{Strange2017Tracking}. TUFF uses open-source materials---including
media reports, official statements, recipient-country systems, and
research outputs---to reconstruct project-level information in the
absence of official Chinese reporting \citep{Strange2017Tracking}.

Both datasets provide project-level information on sector, timing,
geographic coordinates, implementation status, and location precision.
These precision codes are central for empirical design because they
determine how credibly project exposure can be linked to local
outcomes. Precision 1 corresponds to exact project coordinates.
Precision 2 indicates a nearby or buffered location. Precision 3
corresponds to coarser subnational geocoding, often at or through
administrative units or administrative centroids. In the baseline
design, precision 3 observations are retained and mapped through an
ADM2-based rule rather than through the same local-distance rule used
for precision 1.

We apply four sample restrictions. First, we retain only projects whose
start year falls between 2002 and 2013. Because project end dates are
frequently missing---especially in the Chinese data---the start year is
the most consistently observed temporal anchor, and it is the basis for
staggered treatment timing in the panel
\citep{AidData2017WBGeocoded, Strange2017Tracking}. Second, we restrict
the sample to projects classified as ODA-type or otherwise comparable
development assistance, excluding clearly commercial or non-development
financial flows. Third, we retain only projects in implementation or
completed stages, excluding planned or cancelled projects that did not
generate credible local exposure. Fourth, we limit the baseline sample
to projects with precision codes 1--3.

This last restriction deserves emphasis. Retaining precision 1--3
preserves meaningful local variation and maintains comparability with
the panel data used in the analysis. At the same time,
precision 3 is materially coarser than exact coordinates and therefore
introduces the possibility of exposure misclassification. For that
reason, any ``local'' effect estimated from the baseline design should
be interpreted conservatively: it is a local-effect estimate based on
the \emph{reported} project location in the geocoded data, not a claim
that every retained project is known with exact local precision. In
practice, this type of spatial uncertainty is more likely to attenuate
than to spuriously inflate treatment effects, because it adds noise to
the treatment indicator. To make this explicit, treatment assignment in
the baseline is precision-specific: precision 1 uses exact local
matching, precision 2 uses a broader proximity rule, and precision 3
uses an ADM2 rule.

Projects are classified into sectors using the OECD Development
Assistance Committee (DAC) purpose code taxonomy, aggregated to the
three-digit parent-sector level. If a project is coded to multiple
parent sectors, it enters each relevant donor--sector panel separately.
This sector-specific treatment construction is necessary because the
paper's estimands are donor--sector specific rather than pooled across
all forms of aid.

\subsection{Treatment assignment and exposure}\label{assignment}

Our baseline treatment definition is precision-specific rather than a
single uniform distance rule. Throughout, $k$ indexes a donor--sector
panel. For each panel, let $P_k$ denote its set of projects, and for each $p\in P_k$, let
$c_p\in\{1,2,3\}$ denote its precision code and define a match
indicator:
\[
\mathcal{M}_{ip}=\ind\left\{
\begin{aligned}
&c_p=1\ \wedge\ d(i,p)\leq 5\text{ km},\\
&\text{or } c_p=2\ \wedge\ d(i,p)\leq \tau_2,\\
&\text{or } c_p=3\ \wedge\ \mathrm{ADM2}(i)=\mathrm{ADM2}(p)
\end{aligned}
\right\}.
\]
where $\tau_2$ denotes the broader proximity threshold used for
precision-2 records in the dataset. Here, $\tau_2 = 25$ km. This threshold was chosen because
precision-2 records are approximate (``nearby'' or buffered), so a
broader radius is needed to preserve the intended precision hierarchy:
precision 1 captures exact matches, precision 2 nearby matches, and
precision 3 ADM2-level matches. This choice is also consistent with
the spatial support of the outcome data: the coarser raster is
approximately $13.4 \times 13.4$ km, and cluster assignment is based
on the mean of the four nearest 6.7 km cells around the cluster
centroid. Cluster $i$ is then
treated in period $t$ if at least one matching project has started in
period $t$ or earlier:
\[
D_{it}^{k}=\ind\left\{\exists\, p \in P_k \text{ such that }
\mathcal{M}_{ip}=1 \text{ and } s_p \leq t\right\},
\]
where $d(i,p)$ denotes the distance between cluster $i$ and the
reported project location $p$, and $s_p$ is the project's start period.

\begin{figure}[htbp]
\centering
\resizebox{\linewidth}{!}{%
\begin{tikzpicture}[
    font=\small,
    >=Latex,
    box/.style={
        draw,
        rounded corners=2pt,
        thick,
        align=center,
        minimum height=1.0cm,
        inner sep=5pt
    },
    smallbox/.style={
        draw,
        rounded corners=2pt,
        align=center,
        inner sep=4pt
    },
    arrow/.style={->, thick},
    gridcell/.style={draw, minimum width=0.34cm, minimum height=0.34cm, inner sep=0pt},
    title/.style={font=\bfseries\scriptsize, align=center},
    note/.style={font=\scriptsize, align=center}
]

% -------------------------------------------------
% Panel titles
% -------------------------------------------------
\node[title] at (0,3.0) {1. Data inputs};
\node[title] at (4.0,3.0) {2. Treatment mapping};
\node[title] at (8.2,3.0) {3. Outcome surface};
\node[title] at (12.4,3.0) {4. Coarser support};

% -------------------------------------------------
% Panel 1: data inputs
% -------------------------------------------------
\node[smallbox, minimum width=2.85cm, minimum height=2.2cm] (inputs) at (0,1.35) {};

\node[anchor=west, font=\scriptsize] at ($(inputs.west)+(0.18,0.72)$) {$\bullet$ Country};
\node[anchor=west, font=\scriptsize] at ($(inputs.west)+(0.18,0.36)$) {$\bullet$ ADM1};
\node[anchor=west, font=\scriptsize] at ($(inputs.west)+(0.18,0.00)$) {$\bullet$ ADM2};
\node[anchor=west, font=\scriptsize] at ($(inputs.west)+(0.18,-0.36)$) {$\bullet$ DHS cluster};
\node[anchor=west, font=\scriptsize] at ($(inputs.west)+(0.18,-0.72)$) {$\bullet$ Project location};

\node[note, text width=2.6cm] at (0,-0.45)
{Tabular covariates\\and geocoded aid};

% -------------------------------------------------
% Arrow 1
% -------------------------------------------------
\draw[arrow] (1.55,1.35) -- (2.50,1.35);

% -------------------------------------------------
% Panel 2: treatment mapping
% -------------------------------------------------
\node[draw, thick, minimum width=3.3cm, minimum height=2.9cm] at (4.0,1.35) {};

% ADM2 polygon
\draw[thick] (3.45,0.75) -- (3.75,1.95) -- (4.65,2.10) -- (4.85,1.15) -- (4.35,0.62) -- cycle;
\node[font=\scriptsize] at (4.08,1.42) {ADM2};

% Precision 2 radius
\draw[dashed, thick] (4.0,1.35) circle (1.05cm);
\node[font=\tiny, align=center, fill=white, inner sep=1pt] at (4.0,2.40) {Precision 2\\proximity};

% Precision points
\fill (4.62,1.48) circle (2pt);
\node[font=\scriptsize, anchor=west] at (4.72,1.56) {P1};

\fill (3.10,1.35) circle (2pt);
\node[font=\scriptsize, anchor=south east] at (3.00,1.38) {P2};

\fill (4.0,0.45) circle (2pt);
\node[font=\scriptsize, anchor=north] at (4.0,0.34) {P3};

% Radius label
\draw[dashed] (3.10,1.35) -- (4.0,1.35);
\node[font=\scriptsize, above] at (3.55,1.35) {$\tau_2$};

\node[note, text width=4.6cm] at (4.0,-0.50)
{$D_{it}^{k}=1$ if a qualifying project\\matches cluster $i$};

% -------------------------------------------------
% Arrow 2
% -------------------------------------------------
\draw[arrow] (5.65,1.35) -- (6.75,1.35);

% -------------------------------------------------
% Panel 3: outcome grid
% -------------------------------------------------
\begin{scope}[shift={(7.25,0.55)}]
    \foreach \x in {0,...,5} {
        \foreach \y in {0,...,4} {
            \pgfmathtruncatemacro{\shade}{10 + 8*\x + 6*\y}
            \node[gridcell, fill=black!\shade] at (\x*0.34,\y*0.34) {};
        }
    }
    \fill (3*0.34,2*0.34) circle (2.2pt);
\end{scope}

\node[font=\scriptsize, align=center] at (8.1,2.45)
{$Y_{it}\in[0,100]$\\IWI outcome};

% simple legend
\draw[fill=white] (7.35,0.22) rectangle (7.65,0.35);
\draw[fill=gray!35] (7.65,0.22) rectangle (7.95,0.35);
\draw[fill=gray!65] (7.95,0.22) rectangle (8.25,0.35);
\node[font=\scriptsize] at (7.35,0.02) {0};
\node[font=\scriptsize] at (8.25,0.02) {100};

\node[note, text width=3.1cm] at (8.1,-0.45)
{Satellite-imputed\\wealth surface};

% -------------------------------------------------
% Arrow 3
% -------------------------------------------------
\draw[arrow] (9.65,1.35) -- (10.75,1.35);

% -------------------------------------------------
% Panel 4: coarser support
% -------------------------------------------------
\begin{scope}[shift={(11.15,0.82)}]
    \foreach \x in {0,...,5} {
        \foreach \y in {0,...,5} {
            \node[gridcell, minimum width=0.26cm, minimum height=0.26cm] at (\x*0.26,\y*0.26) {};
        }
    }
    \foreach \x in {2,3} {
        \foreach \y in {2,3} {
            \node[gridcell, minimum width=0.26cm, minimum height=0.26cm, fill=gray!35] at (\x*0.26,\y*0.26) {};
        }
    }
    \fill (2.5*0.26,2.5*0.26) circle (1.8pt);
\end{scope}

% zoom box
\draw[dashed, thick] (12.2,1.95) rectangle (13.35,2.75);
\draw[thick] (12.35,2.10) rectangle (13.20,2.60);
\draw (12.775,2.10) -- (12.775,2.60);
\draw (12.35,2.35) -- (13.20,2.35);
\fill (12.775,2.35) circle (2pt);

\draw[dashed] (12.25,1.95) -- (11.90,1.55);
\draw[dashed] (13.30,1.95) -- (12.75,1.55);

\node[note, text width=3.4cm] at (12.35,-0.45)
{Coarser-raster value assigned\\using the documented cluster rule};

% -------------------------------------------------
% Bottom process labels
% -------------------------------------------------
\node[box, minimum width=2.4cm] at (0,-1.85) {Inputs};
\node[box, minimum width=2.8cm] at (4.0,-1.85) {Treatment, $t$};
\node[box, minimum width=2.8cm] at (8.1,-1.85) {Outcome};
\node[box, minimum width=2.2cm] at (12.35,-1.85) {Coarser-raster};

\end{tikzpicture}%
}

\caption{Treatment-definition alternatives and outcome-support mapping.
The schematic summarizes how geocoded aid projects are linked to DHS
clusters and to the satellite-imputed IWI outcome surface. Precision-1
locations are treated as exact local matches, precision-2 locations are
matched using the broader proximity threshold $\tau_2$, and precision-3
locations are assigned at the ADM2 level. The right-hand panel illustrates
the coarser-support treatment-definition alternative, in which outcome
support is aggregated before cluster assignment.}
\label{fig:precision_mapping}
\end{figure}
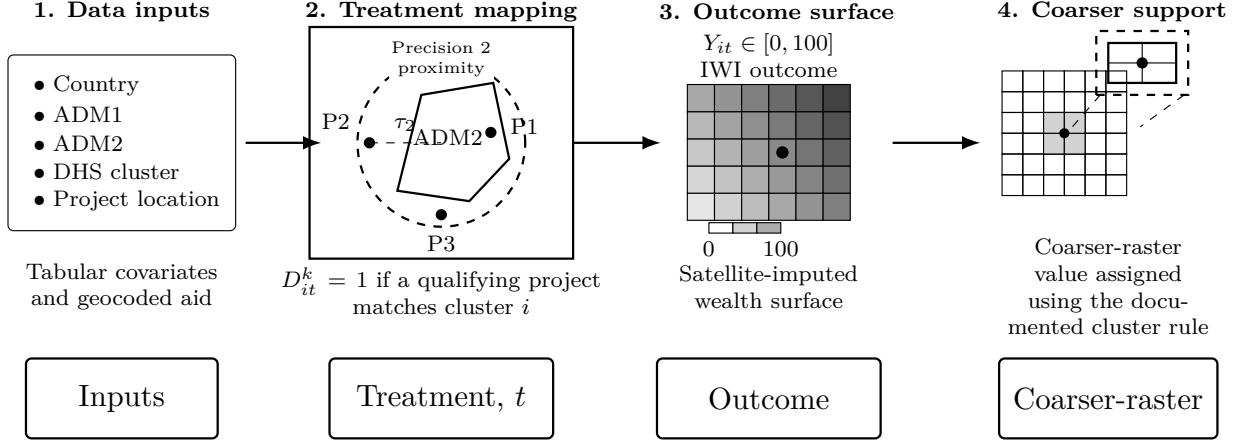

The 5 km local threshold for precision 1 is motivated by the structure
of the DHS data and by the scale of many aid interventions. DHS
coordinates are randomly displaced for confidentiality---up to 2 km in
urban areas and up to 5 km in rural areas for nearly all rural
clusters---so a substantially smaller local radius would amplify spatial
measurement error in matching clusters to projects
\citep{Pettersson2023TimeseriesImagery, Grace2019DHSRemoteSensing}. The
broader precision-2 rule and the precision-3 ADM2 rule reflect the
coarser geocoding information available in those records and preserve
the intended precision hierarchy in the baseline treatment mapping.

Treatment is \emph{absorbing} (i.e., once a cluster enters treatment,
it remains treated in all subsequent periods). This is appropriate for the current
application because projects are dated by start rather than completion,
and because the aim is to capture the onset of persistent project
presence rather than short-lived binary shocks. The resulting treatment
structure is staggered, with different clusters entering treatment in
different periods.

This assumption buys a transparent and consistently observable treatment
clock, but it can obscure several aid-specific timing problems. Some
projects may be short-lived, interrupted, scaled back, or effectively
discontinued after their recorded start date; in that case the
absorbing indicator overstates later exposure and adds measurement error
that would tend to attenuate estimates toward zero. Other projects,
especially roads, dams, power plants, and other infrastructure, may not
become household-relevant until completion, years after recorded start.

Start-year treatment then assigns exposure before the economically
relevant channel is active, which could potentially attenuate early post-treatment coefficients and make the event-time profile look more delayed. A
related problem is cancellation or de facto abandonment after
implementation has begun, which is not fully solved by excluding
projects recorded as planned or cancelled at the outset. Finally, some
projects may generate a temporary local dose that fades after completion
or withdrawal; an absorbing design rules out that reversal by
construction. These concerns are especially relevant for the Chinese
data, where the TUFF methodology reconstructs project information from
open-source materials and completion status is often less consistently
observed.

We nevertheless use start-year absorbing treatment because the main
alternative would require imputing project end dates and operational
dates that are often missing and not consistently comparable across
donors, sectors, or sources. A non-absorbing indicator would therefore
replace one transparent assumption with a harder-to-validate temporal
imputation. The estimates should be read as effects associated with the
onset of recorded local project presence, not as effects of completed or
continuously operating assets in every later period.

Because treatment mapping is a core identifying link in this paper, we
treat two robustness checks as primary design checks: (i)
\emph{strict precision}, which excludes precision-3/ADM2 matches, and
(ii) \emph{coarser outcome raster}, which re-estimates models with a
lower-resolution version of the IWI outcome surface
\citep{daoud2026chinese}. We describe these checks in
Section~\ref{rob-treatment}. A schematic of this precision hierarchy is
reported in Figure~\ref{fig:precision_mapping}.

\subsection{Contextual covariates and overlapping interventions}\label{covariates}

The panel data include several time-varying covariates used in
covariate-adjusted specifications and sensitivity analyses. The main
tabular covariates are log average population density, log conflict
deaths in the prior three years, log disaster counts, an election-year
indicator, a political stability indicator, and a leader-birthplace
indicator. Together, these variables capture urbanization, insecurity,
exogenous shocks, political cycles, institutional stability, and one
important channel of geographically targeted political favoritism.

The data also include variables designed to capture overlapping
interventions and bundled aid exposure. In particular, the data record
logged counts of projects from the other donor, projects in other
sectors from the same donor, and Chinese loan projects. These measures
matter because local project attribution is inherently complicated by
simultaneous or near-simultaneous interventions. A cluster classified as treated in a
given donor--sector panel may also be exposed to other World Bank
sectors, Chinese sectors, or Chinese loan-financed activity. We do not
claim that the baseline design fully eliminates this concern. Rather, we
make it explicit in the data structure and use these variables to inform
sensitivity analysis and interpretation.

This is an important advantage of the panel architecture. Even
without redesigning the main estimator around multi-valued treatment,
the paper can acknowledge that co-treatment is a real feature of the aid
environment rather than treating each donor--sector exposure as if it
occurred in isolation.

\subsection{Donor--sector panels}\label{panels}

The analysis is conducted separately for 23 donor--sector panels. This
design choice is motivated by both substantive and empirical
considerations. Substantively, World Bank and Chinese projects are not
homogeneous interventions and plausibly operate through different
mechanisms and time horizons. Empirically, it
prevents a pooled aid indicator from averaging together sectors whose
targeting patterns, scale, and expected effect timing differ sharply.

Each donor--sector panel is built over the same balanced set of 2,166
DHS clusters observed in four periods, but treatment status varies by
panel depending on whether the cluster is exposed to projects from a
specific donor and sector. The panel composition is summarized below.

\begin{center}
\small
\setlength{\tabcolsep}{8pt}
\begin{tabular}{@{}p{0.44\linewidth}p{0.44\linewidth}@{}}
\toprule
\textbf{China (11 panels)} & \textbf{World Bank (12 panels)} \\
\midrule
Education (110) & Education (110) \\
Health (120) & Health (120) \\
Water Supply \& Sanitation (140) & Water Supply \& Sanitation (140) \\
Government \& Civil Society (150) & Government \& Civil Society (150) \\
Other Social Infrastructure \& Services (160) & Other Social Infrastructure \& Services (160) \\
Transport \& Storage (210) & Transport \& Storage (210) \\
Communications (220) & Communications (220) \\
Energy Generation \& Supply (230) & Energy Generation \& Supply (230) \\
Agriculture, Forestry \& Fishing (310) & Banking \& Financial Services (240) \\
Other Multisector (430) & Agriculture, Forestry \& Fishing (310) \\
Emergency Response (700) & Industry, Mining, Construction (320) \\
 & General Environmental Protection (410) \\
\bottomrule
\end{tabular}
\end{center}

The asymmetry in the number of panels reflects the underlying donor
portfolios and the requirement that each panel contain sufficient
treated and untreated variation to support event-study estimation. 
Some World Bank panels involve comparatively broad spatial coverage,
while several Chinese panels are much sparser. This is not a nuisance
feature of the data; it is part of the empirical environment and affects
the precision with which donor--sector dynamics can be estimated.
Appendix Table~\ref{tab:cohorts} reports balanced-panel accounting, and
Appendix Table~\ref{tab:coverage} summarizes geographic coverage.

\subsection{Empirical strategy and identification}\label{data-identification}

The donor--sector panel structure generates three sources of identifying
variation. First, there is \emph{cross-cluster spatial variation} in
whether a DHS cluster is close to a relevant project location. Second,
there is \emph{temporal variation} in project start timing, because
clusters switch into treatment in different three-year periods. Third,
there is \emph{within-cluster variation}, because the same clusters are
followed over time, allowing fixed effects to absorb time-invariant
differences in baseline geography, survey frame, and other persistent
local characteristics.

A remaining design concern is that period fixed effects absorb only
continent-wide or sample-wide shocks, not shocks that are specific to a
given country in a given period. In this setting, local wealth dynamics
and aid allocation may both respond to country-period factors such as
conflict escalation, election cycles, macroeconomic instability,
commodity-price movements, climatic shocks, or donor-specific shifts in
national engagement. To the extent that such shocks are correlated with
project timing within countries, they may induce residual confounding
that is not removed by cluster and period fixed effects alone.

\paragraph{Benchmark specification: TWFE event study.}
As a benchmark, we estimate a standard staggered-adoption two-way fixed
effects (TWFE) event-study model separately for each donor--sector panel
$k$:
\begin{equation}\label{eq:twfe}
Y_{it}=\alpha_i+\gamma_t+\sum_{\ell \neq 0}
\beta_{\ell}^{\textsc{twfe}}\,
\ind\{t-E_i=\ell\}+\varepsilon_{it},
\end{equation}
where $Y_{it}$ is the IWI of cluster $i$ in period $t$; $\alpha_i$ are
cluster fixed effects; $\gamma_t$ are period fixed effects; and $E_i$
denotes the first treatment period for cluster $i$, with untreated
clusters assigned $E_i=\infty$. The coefficients
$\beta_{\ell}^{\textsc{twfe}}$ trace event time relative to treatment
onset. Under staggered adoption, these TWFE coefficients target a
weighted average of cohort- and period-specific treatment effects rather
than a single clean ATT. In both TWFE and dCdH event-time profiles, event time $\ell=0$
is the omitted reference period, normalized to zero, so
the pre-treatment diagnostic contrast is the last observed untreated-to-onset contrast. Standard errors
are clustered at the DHS-cluster level, the observational unit of the panel.

We treat this specification as a \emph{diagnostic benchmark}. It remains useful because it provides a
familiar reference point for the literature and allows direct comparison
with a preferred switcher--stayer dCdH estimator designed for staggered
treatment timing.

\paragraph{Two distinct identification problems.}
There are two conceptually distinct reasons why TWFE may be misleading
in this setting.

The first is \emph{selective project placement}, which appears as
a non-zero pre-treatment diagnostic contrast. If treated clusters already differ in the contrast between the last untreated period and onset, the parallel-trends assumption needed for
causal interpretation is weakened. In the aid
setting, this is substantively plausible: projects may be directed
toward economically distressed areas, politically salient areas, or
places facing other time-varying shocks. In event-study form, this
problem shows up as a non-zero diagnostic contrast at the available
last untreated-to-onset comparison, which we refer to as the pre-treatment diagnostic contrast.

The second is \emph{staggered-adoption weighting bias}. With absorbing
treatment and heterogeneous effects, the TWFE coefficient is a weighted
average of many implicit two-group/two-period comparisons, and some of
those comparisons can receive negative weights
\citep{deChaisemartin2020TWFE, Callaway2021DiD}. In practice, this means
that already-treated clusters can enter the estimating equation as
controls for later-treated clusters, distorting both static and dynamic
effect estimates.

Negative weighting is mechanical and can arise even under otherwise
acceptable design conditions. Non-zero pre-treatment diagnostic contrasts reflect possible selective
timing or placement and are a deeper threat to identification because
they suggest that switchers and comparison units differ around treatment
onset in ways that may be related to untreated wealth dynamics.

A related but distinct concern is residual country-period confounding.
Even when switchers and comparison units are observed within the same
global period, they may still be exposed to different national shocks if
treatment timing is concentrated in particular countries and years. Our
covariate-adjusted specifications partially address this concern by
conditioning on conflict, disasters, elections, political stability,
and population density, but they do not fully replicate the protection
that would come from saturating the design with country-by-period
effects.

\paragraph{Preferred specification: de Chaisemartin and
d'Haultf\oe{}uille (2020).}
Our preferred switcher--stayer dCdH estimator is the dynamic
difference-in-differences estimator of \citet{deChaisemartin2020TWFE},
applied separately to each donor--sector panel. This estimator avoids the contaminated
early-treated versus late-treated comparisons that drive the negative
weighting problem in TWFE by comparing treatment \emph{switchers} to
same-period \emph{stayers}. Let $\mathcal{S}_t$ denote clusters that
switch from untreated to treated in period $t$. The contemporaneous
effect for switchers can be written as:
\begin{equation}\label{eq:dcdh}
\widehat{\mathrm{ATT}}_0=
\frac{1}{|\mathcal{S}_t|}
\sum_{i\in\mathcal{S}_t}
\left[
(Y_{i,t}-Y_{i,t-1})-
\overline{\Delta Y}_{\text{stayers},t}
\right],
\end{equation}
where $\overline{\Delta Y}_{\text{stayers},t}$ is the mean change in
IWI among clusters whose treatment status does not change in that same
period. Dynamic post-treatment effects are obtained by tracking
switchers forward in event time, and pre-treatment diagnostic contrasts are computed
symmetrically in pre-treatment periods. For direct comparability with
TWFE, we report dCdH event-time coefficients with the same
normalization: $\ell=0$ is the omitted reference period, and the pre-treatment diagnostic contrast is the
reported pre-treatment diagnostic contrast.

To fix ideas, consider a DHS cluster that receives its first World Bank
Health project in the 2005--2007 period. Under TWFE, this cluster can
contribute to event-time comparisons that use clusters treated in
2002--2004 as implicit controls, even though those clusters have
already been exposed to treatment. Under the preferred dCdH
specification, the same switching cluster is compared to clusters that
remain untreated in 2005--2007, the same-period stayers.
The pre-treatment diagnostic contrast is the difference in IWI change between
2002--2004 and 2005--2007 for the switching cluster versus those
stayers. A negative diagnostic contrast means that the switching cluster falls relative
to stayers between the last observed untreated period and treatment onset.
This pattern is consistent with selection bias into weakening areas, but the contrast alone does not identify targeting as the mechanism.

The dCdH estimator is preferred because it is better suited to the
staggered, absorbing-treatment structure of the data.
Three alternative heterogeneity-robust estimators merit discussion.
\citet{Callaway2021DiD} construct group--time average treatment effects
$\text{ATT}(g,t)$ by comparing each treatment cohort to a control group
of either never-treated or not-yet-treated units, then aggregate these
cohort-specific effects into event-study profiles.
This approach is conceptually appealing, but in our setting, it faces a practical
constraint: with merely four three-year periods and 23 donor--sector panels that
vary substantially in the number of treated and untreated clusters, a stable
never-treated control group cannot be maintained uniformly across all panels.
The \emph{not-yet-treated} variant relaxes this requirement, but in a
short staggered panel it progressively depletes the control group as treatment
spreads, leaving later cohorts with few comparison units.

The dCdH switcher--stayer comparison, by contrast, only requires that
some clusters remain untreated within each period; it does not condition on
any particular treatment history prior to that period, making it more robust
to the uneven coverage that characterizes our donor--sector panels
\citep{deChaisemartin2020TWFE}.
The imputation estimator of \citet{Borusyak2024EventStudy} derives
efficient counterfactual predictions from pre-treatment observations and
applies them to the post-treatment periods of switching units.
It achieves efficiency gains under unrestricted treatment-effect heterogeneity,
but its precision depends on the number of pre-treatment periods available
for imputation.
With only a single pre-treatment contrast available in our four-period panel,
the imputation estimator cannot draw on the deeper pre-treatment record that
makes it especially informative; the dCdH first-difference comparison is
more natural for this temporal structure.
Stacked difference-in-differences \citep{Baker2022StackedDiD}
reorganizes the data into cohort-specific sub-experiments and avoids
the contaminated early--late comparisons that bias TWFE.

In practice, however, stacking 23 donor--sector panels---each with its own
treatment cohorts---would multiply the data structure substantially, and the
overlap between cohort windows in a four-period panel makes clean sub-experiment
boundaries difficult to maintain.
The dCdH period-by-period switcher--stayer comparison achieves the same
avoidance of contaminated comparisons without requiring cohort-stacked
datasets, and its cluster bootstrap inference is well-suited to the short
panel and the clustered sampling structure of the DHS data.
Having established why dCdH is the most appropriate estimator for this
setting, it is important to be precise about what it does and does not
solve.
\citet{deChaisemartin2020TWFE} directly addresses the negative-weighting
problem, and by relying on same-period switcher--stayer comparisons it
also removes some contaminated comparisons that can exacerbate bias in
TWFE. But it does \emph{not} automatically restore identification if
switchers and stayers already differ in their untreated trends.
Accordingly, sizable non-zero diagnostic contrasts under the preferred dCdH estimator
will be treated as evidence of residual selection.

\paragraph{Inference and interpretation.}
Inference for the preferred dCdH estimator is based on cluster bootstrap
procedures with 1,000 replications, clustered at the DHS-cluster level.
This accommodates serial dependence within local units over time and is
appropriate for the short panel structure used here
\citep{deChaisemartin2020TWFE}.

The baseline specifications are not covariate-adjusted. This choice is
deliberate: cluster fixed effects absorb time-invariant local
determinants of wealth, period fixed effects absorb shocks common to
all clusters, and the identifying variation comes from differences in
treatment timing rather than from cross-sectional level differences. We
therefore use the unadjusted models as the most transparent baseline
and treat the six-covariate specifications as sensitivity checks for
time-varying shocks that may be correlated with both project timing and
wealth dynamics. As shown in Section~\ref{rob-covariates}, adding these
covariates does not materially change the diagnostic patterns or the
main substantive rankings, which supports the simpler baseline rather
than replacing the core identification strategy.

The interpretation strategy in the results section follows directly from
these diagnostics. Panels with preferred dCdH diagnostic contrasts
close to zero provide the most credible evidence. Panels with modest
remaining diagnostic imbalances are interpreted as suggestive but
informative. Panels with large and statistically clear diagnostic contrasts
are treated with greater caution, even when post-treatment coefficients
are economically interesting. This is the appropriate standard in a
setting where project placement is selective and estimator choice
materially affects the substantive conclusions.

\section{Results}\label{results}

Table~\ref{tab:summary_balance_baseline} summarizes baseline characteristics of the DHS clusters by treatment status across all panels.

% tab_summary_balance_baseline.tex
% Generated from Data/Archive_enriched using baseline period 2002--2004.
\begin{table}[htbp]
  \centering
  \small
  \setlength{\tabcolsep}{4pt}
  \caption{Baseline Summary Statistics and Balance by Treatment Status}
  \label{tab:summary_balance_baseline}
  \begin{tabular}{@{}lcccc@{}}
    \toprule
    & \textbf{All clusters} & \textbf{Ever-treated} & \textbf{Never-treated} & \textbf{Difference (SE)} \\
    \midrule
    IWI & 19.92 (7.84) & 20.30 (8.22) & 19.45 (7.33) & 0.85 (0.33) \\
    Log population density & 4.95 (1.65) & 5.19 (1.59) & 4.66 (1.67) & 0.54 (0.07) \\
    Log conflict deaths & 1.13 (2.17) & 1.24 (2.31) & 1.01 (1.99) & 0.23 (0.09) \\
    Disaster count & 1.38 (5.03) & 1.39 (4.63) & 1.37 (5.46) & 0.02 (0.22) \\
    Election year & 60.1\% & 62.4\% & 57.3\% & 5.08 (2.12) \\
    Political stability & -0.72 (0.75) & -0.72 (0.80) & -0.72 (0.69) & 0.01 (0.03) \\
    \midrule
    Observations & 2,166 & 1,184 & 982 & \\
    \bottomrule
  \end{tabular}
  \vspace{2pt}
  \begin{minipage}{0.96\linewidth}
  \footnotesize\raggedright\textit{Note:} Baseline period is 2002--2004 for the study population cluster units ($N=2{,}166$). Ever-treated units are those treated in at least one donor--sector panel during the 2002--2013 analysis window; never-treated units are untreated in all donor--sector panels. Continuous-variable columns report mean (SD). Election year is reported as a percentage. The Difference column reports ever-treated minus never-treated means, with conventional standard errors for the difference in means in parentheses.
  \end{minipage}
\end{table}

We organize the results around three questions. First, do treated
clusters show evidence of differential pre-treatment wealth trends?
Second, what is the estimated impact of aid on local wealth (IWI) by
donor and sector? Third, how much do substantive conclusions change
when the benchmark TWFE event-study is replaced by the dCdH estimator?
Throughout, we treat pre-treatment diagnostic contrasts as part of the
results, as in this
setting, estimator choice and identification implications are tightly
connected.

\subsection{Diagnostic evidence from pre-treatment contrasts}\label{pretrends}

With four three-year periods, diagnostic evidence comes
from the single observed pre-treatment diagnostic contrast. We therefore focus on whether
treated and untreated clusters differ in that transition from the last
untreated period to the omitted onset period. Table~\ref{tab:pretrend_summary}
summarizes the pre-treatment diagnostic contrast (with $\ell=0$ as the
omitted reference period) across all 23 donor--sector panels under both
the benchmark TWFE event-study and the preferred dCdH estimator.

% tab_pretrend_summary.tex
\begin{table}[htbp]
  \centering
  \small
  \setlength{\tabcolsep}{5pt}
  \caption{Pre-to-onset diagnostic summary by estimator}
  \label{tab:pretrend_summary}
  \begin{tabular}{@{}lrrrr@{}}
    \toprule
    Estimator & Mean at $\ell=-1$ & SD & Sig.\ share & Panels \\
    \midrule
    TWFE & -0.322 & 1.303 & 56.5\% & 23 \\
    dCdH & -0.552 & 0.581 & 69.6\% & 23 \\
    \bottomrule
  \end{tabular}
  \vspace{2pt}
  \begin{minipage}{0.92\linewidth}
  \footnotesize\raggedright\textit{Note:} Summary across all donor--sector panels ($n=23$). Event time $\ell=0$ is the omitted reference period in both estimators. ``Sig.\ share'' is the fraction of panels where the last pre-to-onset contrast at $\ell=-1$ is significant at the 5\% level.
  \end{minipage}
\end{table}

Two patterns stand out. First, non-zero pre-treatment diagnostic contrasts are
common. Under TWFE, the mean diagnostic
coefficient is $-0.32$ IWI points (SD $=1.30$), and 13 of 23 panels
(56.5\%) show statistically significant non-zero contrasts. Under dCdH, the
mean diagnostic contrast is more negative, $-0.55$ IWI points (SD $=0.58$), and 16
of 23 panels (69.6\%) exhibit significant non-zero contrasts. As a heuristic benchmark, under the null of a zero pre-treatment diagnostic contrast
one would expect approximately $23\times 0.05=1.15$ false rejections at
the 5\% level, far below the 13 and 16 observed here. This diagnostic failure is pervasive rather than an isolated anomaly, necessitating a cautious interpretation of the panel-specific estimates

Second, the sign of the contrasts is predominantly negative. Under the preferred dCdH estimator, 15 of the 16 significant diagnostic contrasts are negative. This indicates that, in many panels, clusters that later receive projects were already at weaker relative levels in the pre-treatment diagnostic contrast. We interpret this pattern as evidence consistent with selection bias into economically weaker localities. Importantly, this does not imply that every negative contrast has the same causal origin; it does imply that conventional staggered-TWFE estimates are especially vulnerable in this application because they can attribute recovery from a pre-existing trough to the project itself.

The sector pattern also has a substantive interpretation. Negative
diagnostic contrasts are
consistent with the allocation logics of particular interventions. For
example, Chinese energy projects may be placed in poorer,
resource-adjacent, or infrastructure-deficient areas where new
generation or transmission capacity unlocks extraction, industrial use,
or corridor development rather than immediate household asset
accumulation. Emergency Response is an even clearer case: projects in
that sector are expected to follow conflict, floods, disease outbreaks,
or other acute shocks, which would naturally produce negative
pre-treatment wealth dynamics. By contrast, the comparatively small
preferred dCdH diagnostic contrast for World Bank Health is consistent
with a sector in which poverty targeting and geographic equity mandates
may be more regularized than in infrastructure sectors, where access
costs, construction feasibility, and network placement constraints play
a larger role.

The severity of pre-treatment diagnostic imbalance differs across
panels. Under the TWFE benchmark, the largest negative diagnostic
coefficients for China appear in
Emergency Response ($-3.00$, CI $[-4.48,-1.52]$), Energy Generation and
Supply ($-2.93$, CI $[-3.86,-2.01]$), Other Social Infrastructure and
Services ($-2.34$, CI $[-3.25,-1.43]$), and Water Supply and
Sanitation ($-1.80$, CI $[-2.38,-1.22]$). For the World Bank, the
largest TWFE diagnostic contrasts are observed in Transport and Storage
($-0.88$, CI $[-1.11,-0.65]$), Banking and Financial Services
($-0.77$, CI $[-1.22,-0.32]$), General Environmental Protection
($-0.56$, CI $[-1.07,-0.06]$), Water Supply and Sanitation
($-0.47$, CI $[-0.68,-0.27]$), and Agriculture, Forestry and Fishing
($-0.42$, CI $[-0.63,-0.20]$). The clearest positive TWFE diagnostic contrast is
World Bank Communications ($+2.44$, CI $[1.59,3.28]$), followed by
China Communications ($+1.66$, CI $[0.70,2.62]$) and China Health
($+1.15$, CI $[0.69,1.60]$). These positive contrasts imply a different
distortion: TWFE may produce overly negative post-treatment estimates
when treated clusters already sit above controls in the pre-treatment diagnostic contrast.

The preferred dCdH diagnostic contrasts reinforce rather than remove
the need for caution. Several panels that are substantively important
still display residual selection concerns under dCdH. For China, these
include Education ($-0.969$), Health ($-1.076$), Other Multisector
($-0.752$), Other Social Infrastructure and Services ($-2.342$), and
Water Supply and Sanitation ($-0.737$). For the World Bank, residual
diagnostic imbalance remains in Agriculture ($-0.220$), Education
($-0.684$), Energy ($-0.595$), General Environmental Protection
($-0.486$), Other Social Infrastructure and Services ($-0.703$),
Transport ($-0.485$), and Water Supply and Sanitation ($-0.506$), while
World Bank Communications is the one panel with a positive and
significant dCdH contrast ($+0.332$). By contrast, World Bank Health,
World Bank Government and Civil Society, World Bank Industry, China
Communications, China Energy, China Transport, and China Agriculture
have diagnostic contrasts closer to zero under the preferred dCdH estimator.

These diagnostics motivate the interpretation used below. We
treat panels with small preferred dCdH diagnostic contrasts as the
most credible evidence. Panels with economically important post-treatment
coefficients but significant dCdH diagnostic contrasts are treated as
\emph{suggestive evidence estimated under residual selection concerns}.
This distinction is central to the reading of the results and is
especially important in a paper where the benchmark and preferred
estimators can yield very different substantive conclusions.

\begin{figure}[htbp]
\centering
\includegraphics[width=0.92\linewidth]{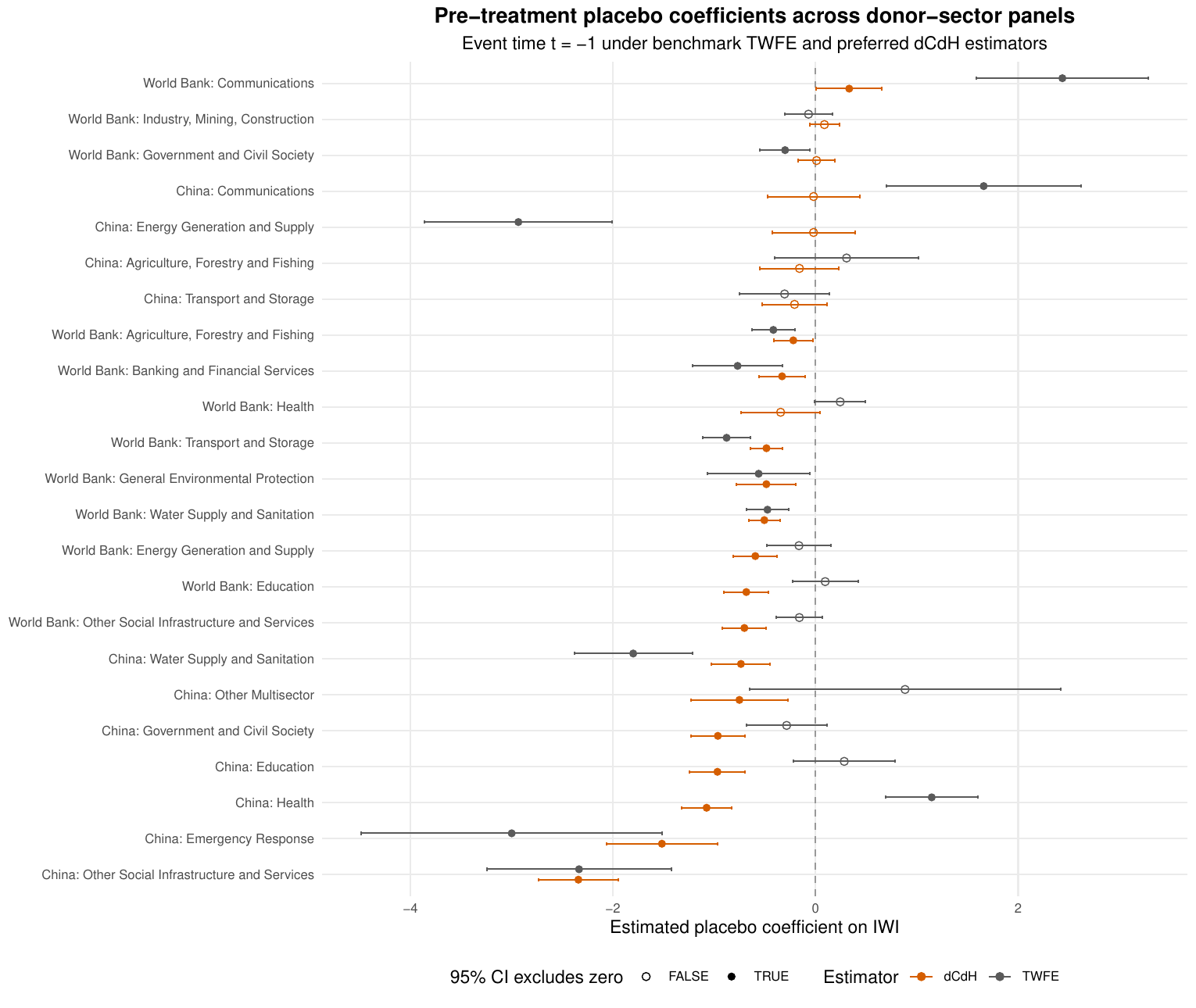}
\caption{Last pre-treatment diagnostic contrasts, reported as the pre-treatment diagnostic contrast (with
$\ell=0$ as the omitted reference period), for all 23 donor--sector panels
under both the benchmark TWFE event-study and the preferred dCdH
estimator. Points show coefficient estimates and bars show 95\%
confidence intervals. Negative values indicate that treated clusters
were below controls in the contrast between the last untreated period and onset. The
prevalence of non-zero contrasts is one reason we place more weight on
the preferred dCdH estimator and on recurring event-time patterns than on
large TWFE coefficients viewed in isolation. Detailed diagnostic estimates
are provided in Appendix
Tables~\ref{tab:pretrend_twfe} and~\ref{tab:pretrend_dcdh}.}
\label{fig:pretrends}
\end{figure}

\subsection{Selected donor--sector patterns under the preferred dCdH estimator}\label{selected-results}

We organize the sector-level discussion around six high-information panels
selected to represent the paper's main empirical contrasts rather than to
exhaust the full set of estimates. The selected panels satisfy at least one of
three criteria: (i) they are substantively central to the donor models discussed
in Section~\ref{background}, especially social-sector delivery for the World Bank and
infrastructure or utility provision for China; (ii) they exhibit large
differences between TWFE and dCdH estimates; or (iii) they illustrate how the
same preferred estimator can yield positive, null, and negative medium-run
profiles. This choice keeps the main text focused while Table~\ref{tab:main_results_summary}
and Appendix Figures~\ref{fig:twfe} and~\ref{fig:dcdh} report the complete 23-panel evidence. The six focal panels
should therefore be read as structured examples that organize the sector-level
interpretation, not as a separate selected sample from which the remaining
panels are excluded.

\paragraph{World Bank: Health, Education, and Agriculture.}
Among World Bank panels, Health provides one of the clearest positive
patterns in the paper. Under the preferred switcher--stayer dCdH estimator, the diagnostic
contrast is small and not statistically distinguishable from zero
($-0.345$, CI $[-0.73,0.04]$), and the dynamic profile shows little
movement immediately after project onset before rising sharply to
$4.42$ IWI points at $\ell=+3$ (CI $[2.26,6.57]$). This delayed profile is
consistent with the idea that household-level asset gains from health
investments do not appear instantly. Substantively, the more important
result is that the preferred dCdH estimate is much larger than the benchmark
TWFE estimate of $1.68$, indicating that the conventional staggered
event-study materially understates the gains in this panel.

World Bank Education also shows a positive preferred dCdH profile,
reaching $3.69$ IWI points at $\ell=+3$ (CI $[3.11,4.26]$).
However, unlike Health, Education retains a negative and significant
dCdH diagnostic contrast ($-0.684$, CI $[-0.91,-0.46]$). We therefore interpret
Education as \emph{suggestive evidence consistent with positive
within-cluster wealth gains, estimated under residual selection
concerns}. The contrast between Health and Education is
useful: both show economically meaningful post-treatment gains, but the
overall design case looks comparatively cleaner for Health than for Education.

World Bank Agriculture moves in the opposite direction. Under the
preferred dCdH estimator, the coefficients are small and mildly positive at
$\ell=+1$ and $\ell=+2$, but the profile turns negative by $\ell=+3$, reaching
$-2.18$ IWI points (CI $[-3.27,-1.09]$). Because the dCdH diagnostic contrast is
also negative and significant ($-0.220$, CI $[-0.41,-0.03]$), we read
this as \emph{suggestive evidence of weak or adverse medium-run wealth
dynamics under residual selection concerns}, not as definitive evidence
of harm. Still, the sign reversal relative to the benchmark TWFE
estimate of $+0.52$ is substantively important.

\paragraph{China: Water Supply and Sanitation, Other Social Infrastructure, Energy, and Transport.}
Among Chinese panels, Water Supply and Sanitation and Other Social
Infrastructure and Services display the two strongest \emph{suggestive}
positive patterns under the preferred dCdH specification. Water Supply and Sanitation reaches
$7.08$ IWI points at $\ell=+3$ (CI $[0.36,13.80]$), while Other Social
Infrastructure and Services reaches $4.44$ (CI $[3.34,5.54]$). These profiles are consistent with positive gains in an asset-based
measure of household well-being, capturing changes in household assets
and living conditions rather than luminosity or market-activity proxies.
However, both estimates are obtained in panels that retain sizable
negative dCdH diagnostic contrasts: $-0.737$ for Water Supply and
Sanitation and $-2.342$ for Other Social Infrastructure and Services.
Accordingly, these should be presented as \emph{suggestive positive
effects estimated under residual selection concerns}, not as the
panels least affected by this diagnostic concern.

China Energy Generation and Supply provides the sharpest example of why
estimator choice is consequential. Under the benchmark TWFE event-study, Energy
appears to generate very large gains, reaching $7.29$ IWI points at
$\ell=+3$ (CI $[4.65,9.93]$). Under the preferred dCdH estimator, the
corresponding estimate is only $0.16$ (CI $[-0.73,1.05]$). This near
collapse of the effect occurs in a panel whose preferred dCdH
diagnostic contrast is essentially zero ($-0.020$, CI $[-0.43,0.39]$), making it
one of the comparatively stronger design cases in the paper. For that reason, China
Energy is especially informative: it shows that a large, apparently
successful TWFE estimate can vanish once contaminated staggered
comparisons are removed.

China Transport and Storage provides a second revealing contrast. The
benchmark TWFE estimate at $\ell=+3$ is positive ($1.04$), whereas the
preferred dCdH estimate turns negative, reaching $-2.42$ IWI points
(CI $[-3.93,-0.91]$) after small or near-zero earlier effects. The
dCdH diagnostic contrast is close to zero ($-0.208$, CI $[-0.53,0.11]$), making
this panel look comparatively cleaner overall than several of the Chinese panels with the largest post-treatment
coefficients. Substantively, the preferred profile is consistent with the
possibility that transport projects generate weak short-run gains but
more disruptive medium-run local reallocation within the observed
window. The relevant reallocation could involve workers and firms moving
toward corridor nodes, roadside markets, or construction-linked
opportunities rather than remaining in the nearby cluster, or road
construction temporarily displacing local farming, housing, or informal
retail activity. We leave mechanism testing for future work, but the empirical
message is already clear: once the staggered-TWFE contamination is
addressed, this sector no longer appears unambiguously beneficial in
household wealth terms.

\begin{figure}[htbp]
\centering
\includegraphics[width=\linewidth]{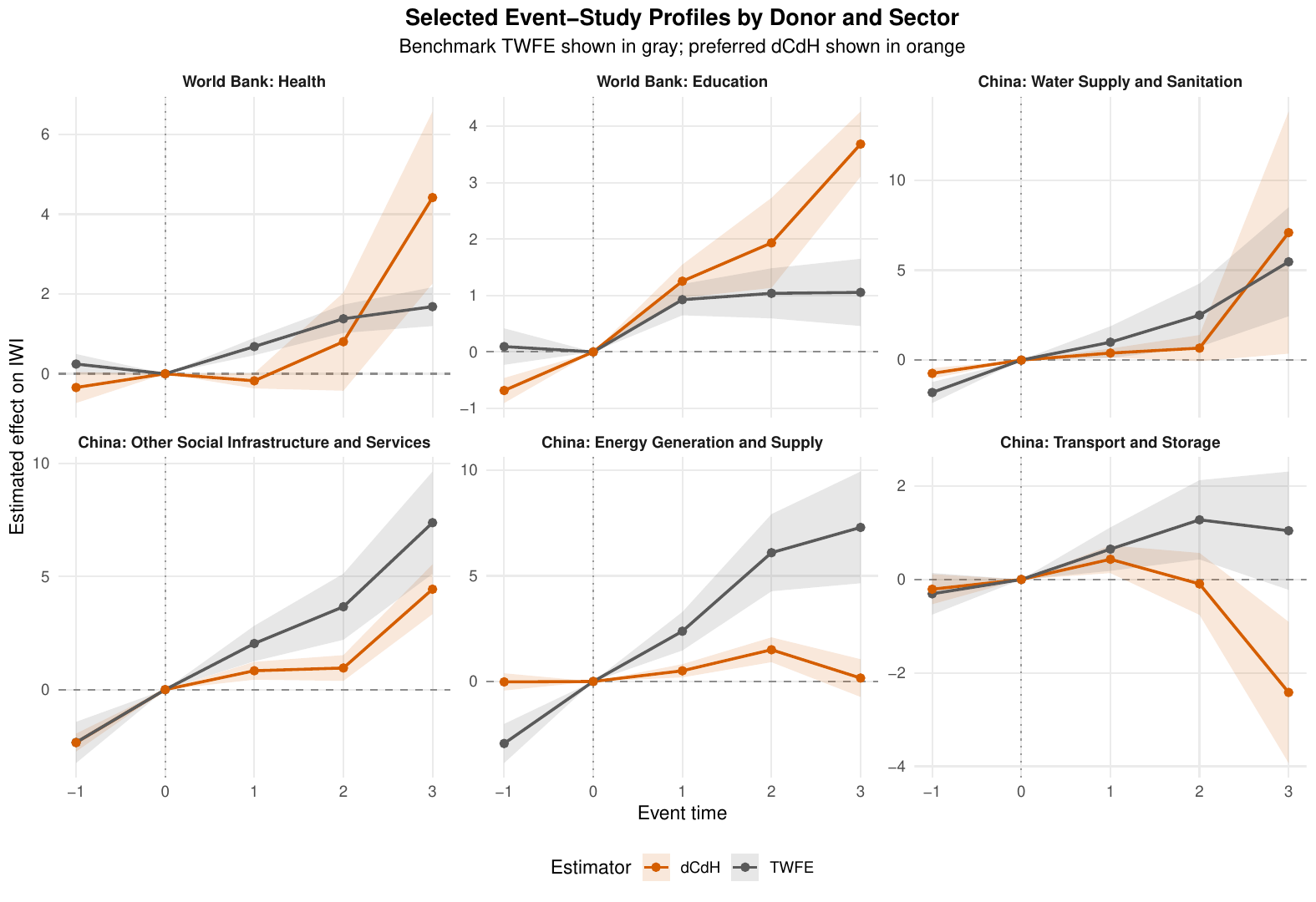}
\caption{Benchmark TWFE and preferred dCdH event-study estimates for six
focal donor--sector panels: World Bank Health, World Bank Education,
China Water Supply and Sanitation, China Other Social Infrastructure
and Services, China Energy Generation and Supply, and China Transport
and Storage. World Bank Health illustrates comparatively cleaner positive evidence;
World Bank Education and the two Chinese social-service panels remain
positive but are estimated under residual selection concerns; China
Energy and China Transport illustrate attenuation and sign reversal once
the benchmark estimator is replaced with the preferred switcher--stayer
design. Both estimators are plotted using the same event-time
normalization, with $\ell=0$ set to zero.}
\label{fig:selected_event_studies}
\end{figure}

\subsection{Estimator choice as a substantive finding}\label{comparison}

The gap between TWFE and dCdH is a central empirical finding,
not an ancillary methodological detail. Across the 23 panels, the preferred dCdH
estimator changes not only magnitudes but often the substantive ranking
of sectors and, in some cases, the sign of the estimated effect. 

To make this comparison concrete, define the estimator gap at the final
event horizon as
\[
\Delta_k=\hat{\beta}^{\textsc{twfe}}_{\ell=+3,k}-
\hat{\beta}^{\text{dCdH}}_{\ell=+3,k}.
\]
Positive values of $\Delta_k$ indicate that TWFE reports a larger
$\ell=+3$ effect than dCdH; negative values indicate that TWFE reports a
smaller effect than dCdH. Because the IWI is measured on a 0--100
scale, gaps of seven to eight points are economically large movements in
the estimator-implied effect, not small numerical differences. We do not
estimate a placebo or bootstrap distribution of TWFE--dCdH gaps, so smaller revisions should be interpreted more cautiously than
the large sign-changing revisions emphasized below.

\begin{figure}[htbp]
\centering
\includegraphics[width=0.92\linewidth]{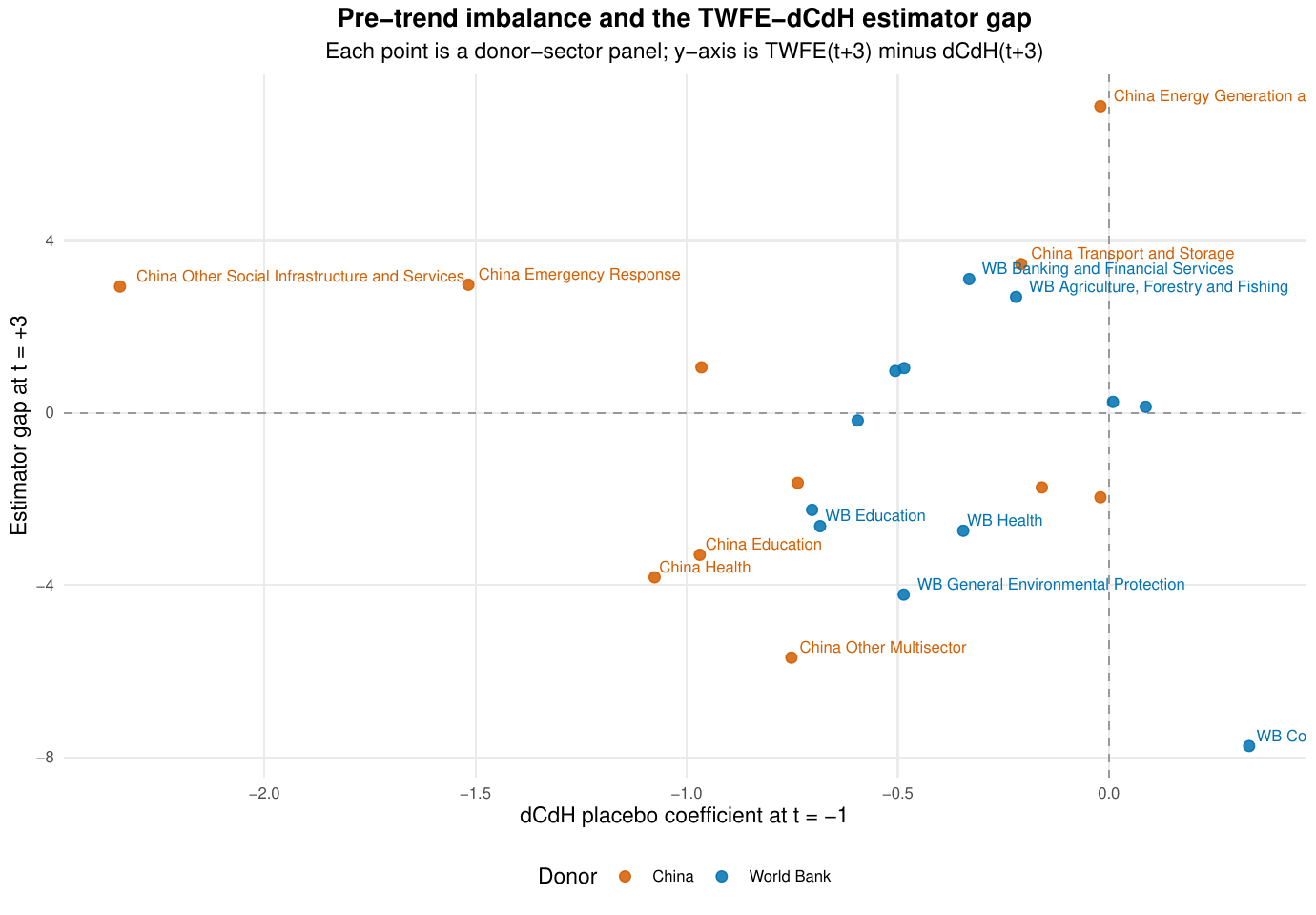}
\caption{Cross-panel relationship between the preferred dCdH
pre-treatment diagnostic contrast (relative to the common $\ell=0$ reference
period) and the estimator gap
$\hat{\beta}^{\textsc{twfe}}_{\ell=+3}-\hat{\beta}^{\text{dCdH}}_{\ell=+3}$.
Panels farther from the vertical zero line exhibit larger diagnostic
imbalances, while panels far from the horizontal zero line exhibit
larger changes in substantive interpretation when TWFE is replaced by
dCdH.}
\label{fig:gap_scatter}
\end{figure}

The largest overstatement in the paper occurs for China Energy, where
the estimate falls from $7.29$ under TWFE to $0.16$ under dCdH
($\Delta=7.13$). China Transport displays another large downward
revision, moving from $+1.04$ to $-2.42$
($\Delta=3.46$). China Emergency Response also attenuates markedly,
from $5.36$ under TWFE to $2.38$ under dCdH. At the same time, several
panels move in the opposite direction. World Bank Health rises from
$1.68$ to $4.42$ ($\Delta=-2.74$), World Bank Education rises from
$1.06$ to $3.69$ ($\Delta=-2.63$), and China Education rises from
essentially zero to $3.30$. The most dramatic reversal of sign appears
in World Bank Communications: the TWFE estimate is $-7.33$ at $\ell=+3$,
but the preferred dCdH estimate is $0.41$, yielding $\Delta=-7.74$.
This panel also has the strongest positive TWFE diagnostic contrast at
the pre-treatment diagnostic contrast ($+2.44$), making it a textbook example of how the benchmark
event-study can produce spurious negative effects when treated clusters
already sit above controls at the pre-treatment diagnostic contrast.
Figure~\ref{fig:gap_scatter} visualizes this comparison across all
panels by plotting the preferred dCdH diagnostic contrast against the
TWFE--dCdH gap at $\ell=+3$.

These examples show that the direction of the correction is closely tied
to the sign of the diagnostic contrast. Panels with negative values of the
pre-treatment diagnostic contrast
tend to see smaller or more negative post-treatment estimates under the
preferred dCdH estimator, consistent with TWFE overstating gains when it
attributes mean reversion to treatment. Panels with positive values of the
pre-treatment diagnostic contrast
tend to move upward under dCdH, because the benchmark event-study had
suppressed post-treatment estimates by comparing treated clusters to an
inappropriately weak counterfactual. Put differently, the same
staggered-TWFE design can bias estimates in either direction depending
on the sign of the pre-treatment diagnostic contrast.

The results support three conclusions. First, patterns consistent with selective project placement are empirically
common and cannot be treated as a secondary robustness issue. Second, once the preferred dCdH estimator is used,
the evidence points to sector-specific rather than donor-wide local
wealth gains. Third, the evidence is concentrated in a subset of panels, especially those in which the preferred dCdH diagnostic contrast is small. Comparatively cleaner panels---those with small pre-treatment diagnostic contrasts---are at least as informative as the largest coefficients, because they identify which donor--sector claims survive both staggered-treatment contamination and residual selection.
This is also why we place more weight on recurring patterns across
estimators and event-time profiles than on isolated significant
coefficients.

\scriptsize\setlength{\tabcolsep}{1.5pt}
\begin{longtable}{@{}>{\raggedright\arraybackslash}p{0.42\linewidth}>{\centering\arraybackslash}p{0.22\linewidth}>{\centering\arraybackslash}p{0.17\linewidth}>{\centering\arraybackslash}p{0.17\linewidth}@{}}
\caption{Main donor--sector summary with confidence intervals and significance notation}\label{tab:main_results_summary} \\
\toprule
\textbf{Sector} & \textbf{Preferred dCdH diag. ($\ell=-1$)} & \textbf{Preferred dCdH ($\ell=+3$)} & \textbf{TWFE ($\ell=+3$)} \\
\midrule
\endfirsthead
\multicolumn{4}{l}{\tablename\ \thetable{} -- continued} \\
\toprule
\textbf{Sector} & \textbf{Preferred dCdH diag. ($\ell=-1$)} & \textbf{Preferred dCdH ($\ell=+3$)} & \textbf{TWFE ($\ell=+3$)} \\
\midrule
\endhead
\midrule
\multicolumn{4}{r}{continued on next page} \\
\endfoot
\bottomrule
\multicolumn{4}{p{\linewidth}}{\footnotesize\textit{Notes:} Summary of the baseline donor--sector comparison. Outcome: IWI points. Unit of observation: DHS cluster-period. Event time $\ell=0$ is the common omitted reference period in both TWFE and dCdH. Asterisks are attached to the preferred dCdH diagnostic contrast at $\ell=-1$ and the preferred dCdH coefficient at $\ell=+3$. Following standard econometric notation, ** indicates that the reported 95\% confidence interval excludes zero (approximately $p<0.05$); no stars indicate that it includes zero. Because this summary table reports 95\% confidence intervals only, * and *** are not used here. Post-treatment columns report the coefficient at $\ell=+3$ with 95\% confidence intervals under the preferred dCdH estimator and TWFE (from Appendix Tables~\ref{tab:a2} and~\ref{tab:a1}).} \\
\endlastfoot
\multicolumn{4}{l}{\textbf{China}} \\
Agriculture, Forestry and Fishing & -0.159 [-0.547, 0.229] & 1.844 [0.68, 3.01]** & 0.116 [-1.63, 1.86]  \\
Communications & -0.020 [-0.475, 0.435] & 2.665 [2.14, 3.19]** & 0.707 [-1.24, 2.65]  \\
Education & -0.969 [-1.243, -0.695]** & 3.302 [2.38, 4.22]** & 0.008 [-1.59, 1.61]  \\
Emergency Response & -1.517 [-2.066, -0.968]** & 2.375 [0.96, 3.79]** & 5.358 [1.94, 8.78]  \\
Energy Generation and Supply & -0.020 [-0.428, 0.387] & 0.162 [-0.73, 1.05] & 7.290 [4.65, 9.93]  \\
Government and Civil Society & -0.965 [-1.230, -0.700]** & -0.321 [-3.25, 2.61] & 0.742 [-0.32, 1.80]  \\
Health & -1.076 [-1.321, -0.831]** & 1.551 [1.03, 2.07]** & -2.265 [-3.41, -1.12]  \\
Other Multisector & -0.752 [-1.230, -0.274]** & 2.228 [1.59, 2.87]** & -3.455 [-6.44, -0.47]  \\
Other Social Infrastructure and Services & -2.342 [-2.736, -1.948]** & 4.442 [3.34, 5.54]** & 7.384 [5.12, 9.64]  \\
Transport and Storage & -0.208 [-0.529, 0.113] & -2.418 [-3.93, -0.91]** & 1.044 [-0.22, 2.31]  \\
Water Supply and Sanitation & -0.737 [-1.027, -0.447]** & 7.082 [0.36, 13.80]** & 5.460 [2.43, 8.49]  \\
\midrule
\multicolumn{4}{l}{\textbf{World Bank}} \\
Agriculture, Forestry and Fishing & -0.220 [-0.414, -0.026]** & -2.179 [-3.27, -1.09]** & 0.521 [-0.01, 1.05]  \\
Banking and Financial Services & -0.331 [-0.556, -0.106]** & -0.689 [-2.16, 0.78] & 2.425 [1.59, 3.26]  \\
Communications & 0.332 [0.009, 0.655]** & 0.412 [-0.04, 0.86] & -7.326 [-8.99, -5.66]  \\
Education & -0.684 [-0.905, -0.463]** & 3.685 [3.11, 4.26]** & 1.055 [0.46, 1.65]  \\
Energy Generation and Supply & -0.595 [-0.811, -0.379]** & 0.276 [-1.03, 1.58] & 0.103 [-0.64, 0.85]  \\
General Environmental Protection & -0.486 [-0.780, -0.192]** & 3.669 [1.75, 5.59]** & -0.549 [-1.87, 0.77]  \\
Government and Civil Society & 0.009 [-0.174, 0.193] & 0.290 [-0.11, 0.69] & 0.549 [0.14, 0.95]  \\
Health & -0.345 [-0.733, 0.043] & 4.418 [2.26, 6.57]** & 1.683 [1.20, 2.17]  \\
Industry, Mining, Construction & 0.087 [-0.060, 0.234] & 0.642 [-0.20, 1.48] & 0.789 [0.18, 1.40]  \\
Other Social Infrastructure and Services & -0.703 [-0.919, -0.487]** & 3.401 [0.66, 6.14]** & 1.151 [0.68, 1.62]  \\
Transport and Storage & -0.485 [-0.645, -0.325]** & 0.776 [0.19, 1.36]** & 1.821 [1.35, 2.29]  \\
Water Supply and Sanitation & -0.506 [-0.658, -0.354]** & 1.309 [0.02, 2.59]** & 2.286 [1.83, 2.75]  \\
\end{longtable}

\normalsize

\section{Robustness}\label{robustness}

This section reports three complementary sets of robustness checks.
First, because treatment mapping is central to identification, we run
\emph{treatment-definition robustness} checks that directly perturb the
mapping from projects to treated clusters and the spatial resolution of
the outcome surface. Second, we report \emph{pooled spatial diagnostics}
designed to assess whether findings could be mechanically driven by
nearby-control contamination or by residual spatial dependence. These
diagnostics are not estimated separately within each donor--sector
panel. Instead, they are estimated on an auxiliary post-treatment
cross-section obtained by stacking the 23 donor--sector panels
and collapsing them to one observation per ISO3--ADM2 unit. Third, we
return to the panel design and re-estimate the benchmark TWFE and
preferred dCdH models with six time-varying covariates already
contained in the panel data. Across these exercises, the basic
substantive picture survives: the most credible positive findings
remain concentrated in a limited subset of donor--sector panels, while
benchmark TWFE positives and panels with sizeable diagnostic
imbalances do not become more persuasive. Figure~\ref{fig:precision_mapping} and Appendix
Table~\ref{tab:app_treatdef_checks} document the treatment-definition
alternatives. Appendix Tables~\ref{tab:a6}, \ref{tab:a7}, and
\ref{tab:a8} report the covariate-sensitivity exercises, and Appendix
Tables~\ref{tab:a3}, \ref{tab:a4}, and \ref{tab:a5} report the
secondary pooled-spatial diagnostics.

\subsection{Treatment-definition robustness}\label{rob-treatment}

The most direct robustness checks for this design are those that alter
the treatment and outcome mapping itself while keeping the estimators,
sample window, and donor--sector panel structure fixed.
Figure~\ref{fig:precision_mapping} documents the
precision-specific mapping, and Appendix
Table~\ref{tab:app_treatdef_checks} documents the two primary
treatment-definition alternatives.

\paragraph{Strict precision mapping.}
In the strict-precision specification, treatment is re-defined by
excluding all precision-3 records and their ADM2-based assignments. In
practice, this means that treatment can only be activated by precision-1
exact local matches and precision-2 broader proximity matches. Formally,
the strict indicator is
\[
D_{it}^{k,\text{strict}}=
\ind\left\{\exists\, p\in P_k:\ c_p\in\{1,2\},
\mathcal{M}_{ip}=1,\ s_p\leq t\right\}.
\]
This check isolates how much of the baseline signal depends on the
coarsest geocoding layer (precision 3/ADM2).

\paragraph{Coarser outcome raster.}
We also define a coarser-raster treatment-mapping exercise using the same TWFE and dCdH specifications but a coarser
version of the IWI outcome raster. This follows the design logic in
\citet{daoud2026chinese}: if estimates are sensitive
to the native 6.72 km support of the imputed outcome, that sensitivity
should be visible when outcomes are aggregated to a lower spatial
resolution before cluster-level assignment. In the appendix materials,
this alternative is documented at approximately $13.4 \times 13.4$ km support, and each DHS cluster is assigned using the mean of the four nearest 6.7 km cells around the cluster centroid.

These treatment-definition checks receive more emphasis
than pooled-ADM2 diagnostics because they address the
project-to-outcome mapping underlying the identifying variation more directly.

These checks vary the spatial mapping of treatment and the outcome
raster, but they do not vary the absorbing temporal structure. That
choice reflects the same data constraint that motivates the baseline
definition: project end dates and operational dates are too incomplete
to support a consistently comparable non-absorbing treatment indicator
across donors and sectors. Violations of the absorbing assumption should
therefore be read as a limitation of the temporal treatment measure.
Short-lived or discontinued projects could potentially attenuate estimates toward zero,
while delayed operationalization could possibly depress early post-treatment coefficients and shift plausible effects toward
later event-time horizons.

\subsection{Secondary pooled spatial diagnostics}\label{rob-spatial}

\paragraph{Construction of the pooled diagnostic sample.}
For the spatial robustness exercises, the 23 donor--sector panels are
stacked and then collapsed to a single post-treatment cross-section.
The collapse is carried out at the ISO3--ADM2 level (second-level administrative units, equivalent to districts or counties) to avoid counting
the same geography multiple times across donor--sector panels. For each
ADM2 unit, we compute average latitude and longitude, average covariate
values, and a pooled ever-treated indicator that equals one if any of
the stacked donor--sector data record treatment prior to the
post-treatment outcome windows. Post-treatment wealth is measured as the
mean of the satellite-imputed IWI values in 2014--2016 and 2017--2019.
This construction yields a broad diagnostic sample suitable for
examining generic spatial spillovers. Its purpose is narrower: to test whether nearby untreated areas
are likely to be contaminated by positive externalities and whether
residual spatial dependence is severe enough to warrant caution in
inference.

\paragraph{Distance-based dose-response.}
If aid projects generate substantial positive spillovers to nearby
untreated areas, one would expect units close to treated sites to
display \emph{higher} post-treatment IWI than units farther away, all
else equal. To test this prediction, we compute the Haversine distance
from each pooled ADM2 unit to the nearest treated unit and partition
space into four concentric bands: $[0,10)$ km, $[10,20)$ km,
$[20,50)$ km, and $[\geq 50)$ km, with the farthest band serving as
the omitted reference group. We then estimate the following auxiliary
OLS model:
\begin{equation}\label{eq:dose}
Y_i^{\text{post}} = \alpha + \delta D_i + \sum_{b \in \mathcal{B}}
\gamma_b \cdot \ind\{i \in b\} + \mathbf{X}_i'\boldsymbol{\beta}
+ \mu_c + \varepsilon_i,
\end{equation}
where $Y_i^{\text{post}}$ is the pooled post-treatment IWI, $D_i$ is
the pooled ever-treated indicator, $\mathcal{B}=\{[0,10), [10,20),
[20,50)\}$ is the set of non-reference distance-band indicators,
$\mathbf{X}_i$ contains log average travel time to the nearest city, log
average population density, and log three-year pre-period conflict
deaths, and $\mu_c$ are country fixed effects. Inference is based on
HC2 heteroskedasticity-robust standard errors \citep{MacKinnon1985HC}.

The distance-band coefficients do not support the positive-spillover
hypothesis. Relative to units at least 50 km away from any treated
location, units within 0--10 km show an IWI deficit of $-2.23$ points
($p = 0.065$), units within 10--20 km show a deficit of $-1.27$
($p = 0.110$), and units within 20--50 km show a deficit of $-2.66$
($p < 0.001$); see Appendix Table~\ref{tab:a3}. The fact that nearby
units are \emph{poorer}, not richer, than distant ones directly
contradicts the idea that the main results are being generated by
positive spillovers from treated locations into nearby controls.
Instead, the pattern is more consistent with geographic targeting to
weaker local economies, such that both treated and nearby untreated
areas are located in generally poorer regions. This exercise therefore
points away from positive spillovers as the main explanation for the
results and toward spatial concentration of need and selective
placement.

\paragraph{Exclusion buffers.}
A stronger version of the spillover concern is that the untreated units
closest to treated sites may receive enough indirect benefit to bias the
control group upward, thereby attenuating the estimated treatment
effect. We evaluate this possibility by progressively excluding control
units located within 10 km, 20 km, and 30 km of any treated site, while
always retaining treated units. For each buffer radius
$K \in \{0,10,20,30\}$ km, we estimate:
\begin{equation}\label{eq:buffer}
Y_i^{\text{post}} = \alpha^{(K)} + \delta^{(K)} D_i +
\mathbf{X}_i'\boldsymbol{\beta}^{(K)} + \mu_c^{(K)} +
\varepsilon_i^{(K)},
\end{equation}
on the restricted sample consisting of all treated units plus untreated
units farther than $K$ kilometers from the nearest treated location.
If nearby-control contamination were important, the treatment
coefficient should increase materially as larger exclusion buffers are
imposed.

The estimates show no such pattern. The baseline pooled treatment
coefficient is $1.395$ IWI points with no exclusion buffer. Excluding
controls within 10 km yields $1.368$; excluding controls within 20 km
yields $1.486$; and excluding controls within 30 km yields $1.195$.
All four estimates remain positive and statistically significant, and
their confidence intervals overlap heavily; see Appendix
Table~\ref{tab:a4} and Appendix Figure~\ref{fig:buffer_robustness}. The
modest decline in the 30 km specification is
more plausibly attributed to loss of precision as the control sample
shrinks than to any systematic shift in the treatment effect. The
buffer exercise therefore suggests that the main positive donor--sector
patterns are unlikely to be artifacts of nearby untreated units being
``too treated'' through local spillovers.

\paragraph{Residual spatial autocorrelation.}
Even if nearby-control contamination is limited, inference may still be
too optimistic if regression residuals are spatially clustered. We test
for this using Global Moran's $I$ on the residuals from the
distance-band model in Equation~\eqref{eq:dose}. Following standard
practice in spatial econometrics \citep{Anselin1988Spatial}, we
construct a five-nearest-neighbor spatial weights matrix and
row-standardize it so that each unit's weights sum to one. The Moran
statistic is:
\begin{equation}\label{eq:moran}
I = \frac{N}{\sum_i \sum_j w_{ij}}
\cdot
\frac{\sum_i \sum_j w_{ij}(e_i-\bar{e})(e_j-\bar{e})}
{\sum_i (e_i-\bar{e})^2},
\end{equation}
where $e_i$ denotes the residual for unit $i$, $\bar{e}$ is the sample
mean residual, and $w_{ij}$ are the elements of the spatial weights
matrix.

The resulting Global Moran's $I$ is $0.239$, compared to an expected
value under spatial independence of approximately $-0.0005$, with
variance $0.000162$ and $p<0.001$; see Appendix
Table~\ref{tab:a5}. A Monte Carlo permutation test with 999
randomizations yields the same qualitative conclusion. Residual spatial
clustering means that nearby clusters still share unobserved conditions
that shape wealth trajectories, such as access to the same markets,
local government capacity, agroclimatic conditions, or exposure to the
same infrastructure network. Economically, the implication is that treated
and nearby untreated clusters may be more comparable than geographically
distant clusters on some dimensions, while still sharing unobserved
shocks that conventional cluster-level inference does not absorb.
Clustered standard errors at the DHS-cluster level are necessary in the
main panel models because they account for serial dependence within
observational units over time \citep{deChaisemartin2020TWFE}; however,
they do not fully eliminate the possibility of cross-unit spatial
dependence. The Moran diagnostic therefore reinforces a conservative
reading of geographically clustered donor--sector estimates.

The pooled spatial diagnostics do not support the view that the main
findings are driven by positive spillovers into nearby controls. If
anything, the generic spatial pattern is the opposite: treated and
nearby untreated locations tend to be poorer than distant ones, which is
more consistent with geographic targeting into weaker
local economies. At the same time, the Moran diagnostic confirms that
residual spatial dependence remains present, so the results should be
interpreted conservatively and not as if spatial independence were
fully resolved.

\subsection{Covariate sensitivity in the panel estimators}\label{rob-covariates}

After the treatment-definition checks above, an additional panel-level
robustness check is to re-estimate the main donor--sector ATT models
after conditioning on observed time-varying factors that plausibly
shape both project placement and local wealth dynamics. These covariates
provide only a partial check against country-period confounding, because
they capture several time-varying shocks and political conditions that
may jointly influence aid timing and local welfare trajectories. We
therefore re-estimate the TWFE and dCdH specifications using six
covariates available in the panel data:
\[
\mathbf{X}_{it} =
\left(
\log \text{PopDens}_{it},
\log \text{Conflict}_{it},
\log \text{Disasters}_{it},
\text{ElectionYear}_{it},
\text{PoliticalStability}_{it},
\text{LeaderBirthplace}_{it}
\right).
\]
These variables are intended to proxy urbanization, local insecurity,
natural shocks, electoral cycles, institutional fragility, and one
important channel of political favoritism in aid placement.

\paragraph{TWFE with covariates.}
The covariate-adjusted TWFE model takes the form
\begin{equation}\label{eq:twfe_cov}
Y_{it} = \alpha_i + \gamma_t + \sum_{\ell \neq 0}
\beta_{\ell}^{\textsc{twfe-cov}}
\cdot \ind\{t-E_i=\ell\}
+ \mathbf{X}_{it}'\boldsymbol{\phi} + \varepsilon_{it}.
\end{equation}
As in the baseline specification, event time $\ell=0$ is the omitted
reference period. Appendix Table~\ref{tab:a6} reports the full
panel-by-panel covariate-adjusted TWFE coefficients. These profiles are
numerically close to the benchmark TWFE estimates, which is
substantively useful even if unsurprising: once unit and period fixed
effects and event-time indicators are in the model, the six covariates
do not materially alter the TWFE profiles or their substantive
ranking. But this stability does not repair the core identification
problem of TWFE in this setting. The main difficulty is not merely
omission of a few tabular covariates; it is the combination of
selection bias and staggered-treatment contamination documented in the
previous section.

\paragraph{dCdH diagnostic contrasts with covariates.}
The more relevant covariate-sensitivity question concerns the preferred
estimator. Appendix Table~\ref{tab:a7} reports the full panel-by-panel
dCdH diagnostic contrasts from the six-covariate specification where
feasible, with the estimator's no-covariate fallback used in four sparse
panels. These diagnostics remain close to the baseline pattern: panels
that look comparatively clean without covariates largely remain the
cleaner panels with covariates, while the main non-zero diagnostic
contrasts also persist. Thus, controlling for the six observed
covariates does not materially change which panels look comparatively
cleaner, which remain suggestive, and which remain exposed to selection
concerns.

\paragraph{dCdH post-treatment estimates with covariates.}
Appendix Table~\ref{tab:a8} is the key lookup table for the
covariate-adjusted dCdH coefficients and reports the full panel-by-panel
estimates, using the estimator's no-covariate fallback in four sparse
panels when the six-covariate specification does not converge. The
central donor--sector patterns largely survive, but their interpretive
strength continues to depend on the diagnostic contrast for each panel.
In substantive terms, covariate adjustment changes some magnitudes and
precision, but it does not overturn the paper's main ranking: the most
robust World Bank gains remain concentrated in Health, while the
largest Chinese positive coefficients remain concentrated in selected
utility and social-infrastructure sectors rather than in Energy,
albeit under continued diagnostic-contrast-based selection concerns.

\paragraph{In sum,} the robustness exercises point in a similar direction. Treatment-definition alternatives are the primary design checks: strict precision and coarser-raster re-mapping directly probe whether the core project-to-outcome linkage is carrying the findings. The pooled spatial diagnostics are secondary and suggest that positive spillovers into nearby controls are unlikely to be the primary explanation for the results, though residual spatial dependence remains present. The covariate- adjusted panel estimators show that the paper's main conclusions are not simply artifacts of omitting a small set of observed political and geographic controls. 

At the same time, these checks do not eliminate the
core identification discipline established earlier: panels with sizeable
dCdH diagnostic contrasts remain panels that should be interpreted as
suggestive. Robustness in this paper therefore
means that the main donor--sector patterns remain similar across several
alternative checks while still being read in light of the
design concerns laid out earlier, especially the remaining possibility
of country-period shocks correlated with treatment timing.

\section{Discussion and limitations}\label{interpretation}

\paragraph{Interpretation.}
The results point to a selective rather than donor-wide account of local
wealth effects. Selection bias is not a peripheral
complication: in many panels, treated clusters were already on weaker
relative levels in the pre-treatment diagnostic contrast, and the benchmark staggered-TWFE estimator
is prone to attribute subsequent catch-up from those imbalances to
treatment. Once estimates are read through the switcher--stayer design and the
pre-treatment diagnostic contrasts, the
World Bank evidence is concentrated in sectors closely tied
to service delivery and human capital, while Chinese
patterns are concentrated in selected utility and social-infrastructure
panels rather than in Energy.

The paper also underscores that the choice of outcome variable shapes
which sector effects are visible. Because the IWI is a household-centered measure of
assets and living conditions, infrastructure-heavy sectors can look
weaker here than in studies based on nighttime lights or other
activity-centered proxies \citep{Jean2016SatelliteML, Bluhm2018Connective, Pettersson2023TimeseriesImagery}.
A near-zero estimate for China Energy should therefore not be read as
evidence that such projects have no economic effects at all. It should
be read more narrowly as evidence that, within the observed
post-treatment window and using an asset-based welfare outcome, those
effects do not translate into robust local wealth gains.

The results further suggest only a cautious form of donor
complementarity. The World Bank appears most favorable in Health and, more
tentatively, Education, whereas China's more favorable patterns appear
in Water Supply and Sanitation and Other Social Infrastructure and
Services. But several panels remain weak or negative, and the
interpretive strength of the estimated effects varies across sectors. A cautious reading is therefore that local wealth gains vary along both donor and sector dimensions.

\paragraph{Limitations.}
Several limitations bound the scope of inference. The outcome is model-generated rather than directly observed in repeated household surveys, so prediction error remains part of the data-generating process \citep{Pettersson2023TimeseriesImagery}. The baseline treatment definition is precision-specific (precision 1 exact local matching, precision 2 broader proximity matching, and precision 3 ADM2 matching), which is transparent and workable but still leaves measurement error in treatment status for coarser geocoding layers. The strict-precision and coarser-raster treatment-definition alternatives directly probe this mapping choice, but they cannot fully eliminate all spatial uncertainty. The design also remains vulnerable to bundled interventions and overlapping treatment: the panel data record other-donor exposure, other-sector exposure, and Chinese loan exposure, yet the main estimator is still organized around donor--sector panels rather than a fully multivalued treatment framework.

In addition, the paper
evaluates 23 donor--sector panels and multiple dynamic coefficients
within each panel, so isolated statistically significant estimates
should not be over-interpreted on their own; more weight should be
placed on recurring patterns across estimators, event-time profiles,
and pre-treatment diagnostic contrasts. Country-specific shocks that
vary over time also remain only partially addressed: cluster and period
fixed effects do not eliminate the possibility that country-period
events --- such as macroeconomic crises, conflict escalation, climatic
shocks, or national political transitions --- are correlated with both
project timing and local wealth dynamics, and the covariate-adjusted
specifications provide only partial protection on that margin. 

The
robustness checks further suggest that positive spillovers into nearby
controls are unlikely to be the main explanation for the results, yet
residual spatial autocorrelation remains present. Finally, the
observation window is long enough to capture medium-run adjustment but
may still be too short for some infrastructure projects whose
household-welfare effects unfold over longer horizons. These
limitations do not undo the paper's main empirical message, but they do
mean that the paper should be read as providing a careful map of
which donor--sector patterns remain credible under the available
design, not as a final accounting of all local effects of aid in
Africa.

\section{Conclusion}\label{conclusion}

This paper studies the local wealth effects of World Bank and Chinese development projects in Africa using a balanced panel of 2,166 DHS clusters across 35 countries and a satellite-imputed measure of household material well-being. The main empirical message is methodological as well as substantive. In this setting, conventional staggered-TWFE event-studies are not neutral summary tools: they are sensitive to selective project placement and to negative-weight contamination under heterogeneous treatment timing. Once those issues are confronted directly, the apparent ranking of sectors changes materially.

The resulting picture is more selective than the benchmark TWFE results suggest. Among the World Bank panels, Health shows positive evidence, with delayed gains in local wealth, while Education is also positive but estimated under residual selection concerns. Among the Chinese panels, Water Supply and Sanitation and Other Social Infrastructure and Services remain positive but should be interpreted as suggestive because of non-zero pre-treatment diagnostic contrasts. China Energy and China Transport are among the informative panels because they look less favorable once the preferred dCdH estimator replaces TWFE: the former collapses to approximately zero and the latter turns negative over the medium run. Overall, the results do not support a donor-wide verdict that either the World Bank or China uniformly ``works.'' The narrower conclusion is that within-neighborhood wealth gains are sector-specific, and the most cautious claims are concentrated in a subset of panels rather than in isolated significant estimates viewed one by one.

Substantively, the positive World Bank Health and Education profiles suggest that social-sector aid can translate into household material gains when projects affect the services, assets, and local conditions that enter everyday welfare, although the evidence is cleaner for Health. For China, the more favorable patterns in Water Supply and Sanitation and Other Social Infrastructure point to locally visible welfare channels in basic services and community infrastructure, while Energy and Transport suggest that large infrastructure portfolios may produce benefits outside the observed neighborhoods or beyond the medium-run window rather than immediate within-neighborhood asset gains.

More broadly, the paper shows what subnational aid evaluation gains from combining improved local outcomes with credible causal design. Satellite-imputed wealth data make it possible to study within-neighborhood change at a scale that traditional survey data alone rarely permit. But better outcome measurement does not substitute for identification. Future research should extend the time horizon, impose stricter geocoding and treatment-overlap designs, and more directly connect dynamic DiD approaches to image-augmented assignment models. Those steps would help clarify which local wealth effects persist over longer horizons, which are specific to the medium-run donor--sector environment studied here, and how much remaining heterogeneity is driven by country-period shocks rather than treatment itself. \hfill $\blacksquare$

\begingroup
\sloppy
\raggedright
\hbadness=10000
\hfuzz=100pt
\bibliographystyle{apalike}
\bibliography{mybib}
\endgroup

\clearpage
\section{Appendix}\label{appendix}
\setcounter{figure}{0}
\setcounter{table}{0}
\setcounter{equation}{0}
\renewcommand{\thefigure}{A.\arabic{figure}}
\renewcommand{\thetable}{A.\arabic{table}}
\renewcommand{\theequation}{A.\arabic{equation}}
\renewcommand{\theHfigure}{A.\arabic{figure}}
\renewcommand{\theHtable}{A.\arabic{table}}
\renewcommand{\theHequation}{A.\arabic{equation}}

\subsection{Primary treatment-definition robustness}

\begin{table}[htbp]
  \centering
  \caption{Treatment-definition alternatives}
  \label{tab:app_treatdef_checks}
  \begin{tabular}{p{0.23\linewidth}p{0.34\linewidth}p{0.37\linewidth}}
    \toprule
    \textbf{Check} & \textbf{Operational change} & \textbf{Design purpose} \\
    \midrule
    Baseline precision hierarchy
      & Precision 1: local 5~km match; Precision 2: broader proximity rule $\tau_2$; Precision 3: ADM2 match
      & Preserve information in precision 1--3 while honoring geocoding-quality differences in treatment mapping. \\
    Strict precision
      & Exclude precision-3 records and all ADM2-based treatment assignments; retain only precision 1--2 mapping
      & Test whether baseline findings depend on the coarsest geocoding layer. \\
    Coarser outcome raster
      & Define the same TWFE and dCdH estimands under lower-resolution IWI support before cluster-level assignment
      & Test sensitivity of estimated effects to the spatial support of the model-generated outcome. \\
    \bottomrule
  \end{tabular}
  \begin{flushleft}
  \footnotesize\textit{Notes:} This table documents the treatment-definition alternatives discussed in Section~\ref{rob-treatment}. For precision-2 locations, $\tau_2=25$ km. The coarser-raster alternative uses outcomes aggregated to approximately $13.4 \times 13.4$ km support, after which each DHS cluster is assigned using the mean of the four nearest 6.7 km cells around the cluster centroid. These appendix materials document the mapping alternatives rather than reporting full re-estimated results.
  \end{flushleft}
\end{table}

\clearpage
\subsection{Main panel estimates and estimator diagnostics}

\begin{figure}[htbp]
\centering
\includegraphics[width=\linewidth]{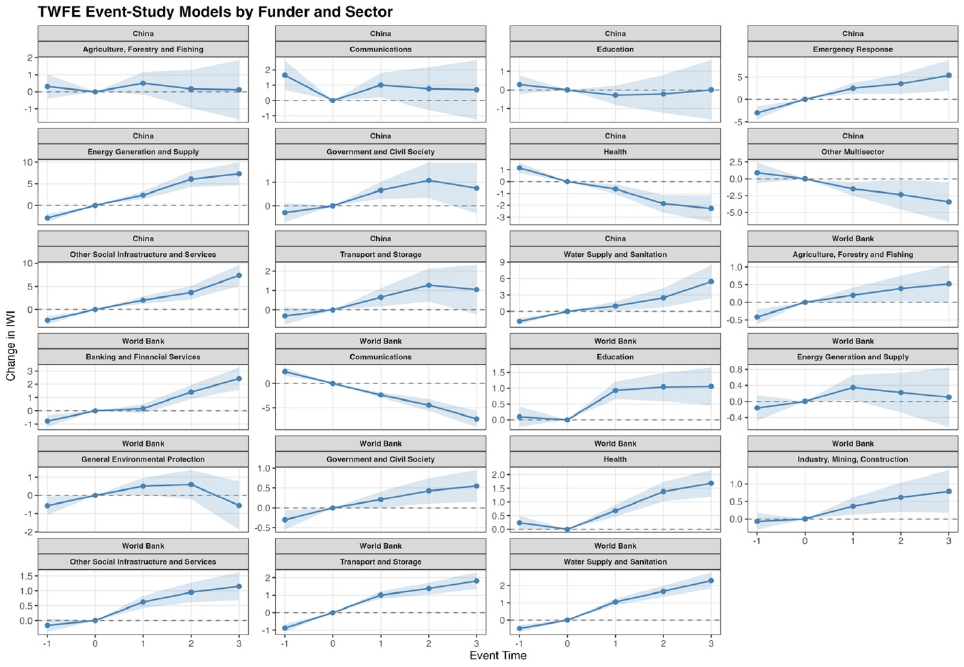}
\caption{Benchmark TWFE event-study estimates across all 23 donor--sector panels, normalized with event time $\ell=0$ as the omitted reference period. These coefficients are useful diagnostically but should not be treated as preferred causal estimates in a setting with selection bias and staggered treatment timing.}
\label{fig:twfe}
\end{figure}

\clearpage
\begin{figure}[htbp]
\centering
\includegraphics[width=\linewidth]{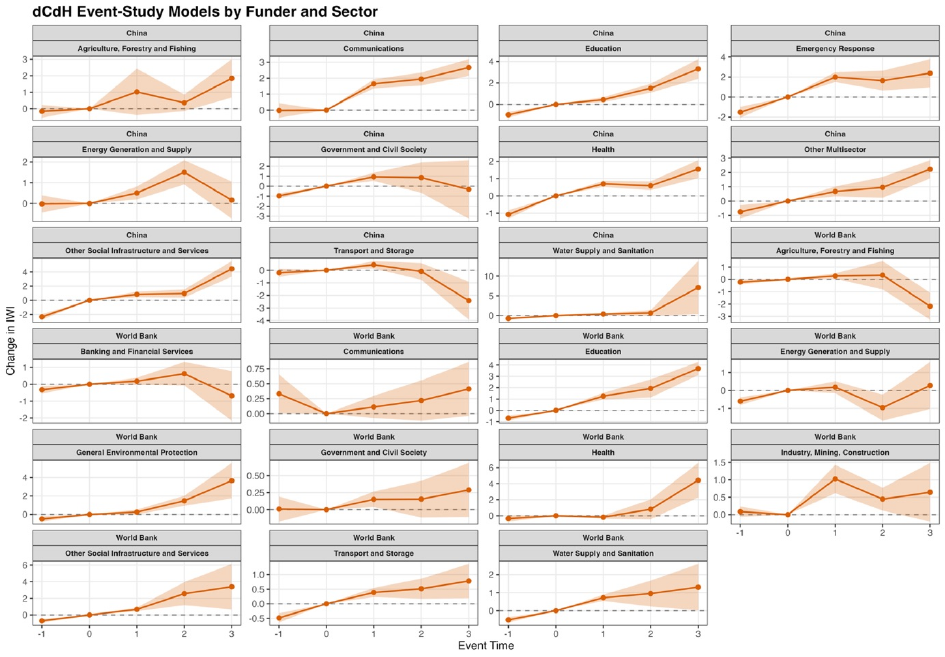}
\caption{Preferred dCdH event-study estimates across all 23 donor--sector panels, reported on the same event-time normalization as TWFE with $\ell=0$ as the omitted reference period. Confidence intervals are obtained via cluster bootstrap. Panels with non-zero diagnostic contrasts remain informative but should be interpreted under residual
selection concerns.}
\label{fig:dcdh}
\end{figure}

\clearpage
% tab_cohorts.tex
\begin{table}[htbp]
  \centering
  \caption{Balanced Panel Accounting for DHS Clusters}
  \label{tab:cohorts}
  \begin{tabular}{lrr}
    \toprule
    Period & Clusters & Observations \\
    \midrule
    2002--2004 & 2,166 & 2,166 \\
    2005--2007 & 2,166 & 2,166 \\
    2008--2010 & 2,166 & 2,166 \\
    2011--2013 & 2,166 & 2,166 \\
    \midrule
    \textbf{Total clusters (N)} & \textbf{2,166} & -- \\
    \textbf{Total observations} & -- & \textbf{8,664} \\
    \bottomrule
  \end{tabular}
  \vspace{4pt}
  \begin{minipage}{0.92\linewidth}
  \footnotesize\raggedright\textit{Note:} Balanced panel with identical spatial units across four periods; total observations equal $2,166 \times 4 = 8,664$.
  \end{minipage}
\end{table}

\clearpage
% tab_coverage.tex
\begin{longtable}{@{}llr@{}}
  \caption{DHS Cluster Coverage by Country\label{tab:coverage}} \\
  \toprule
  Country & ISO3 & DHS Clusters \\
  \midrule
  \endfirsthead
  \multicolumn{3}{c}{\tablename\ \thetable{} (continued)} \\
  \toprule
  Country & ISO3 & DHS Clusters \\
  \midrule
  \endhead
  \endfoot
  \bottomrule
  \multicolumn{3}{l}{\small\textit{Note:} Balanced panel includes 2,166 DHS clusters.} \\
  \endlastfoot
  Nigeria & NGA & 286 \\
  Kenya & KEN & 204 \\
  Malawi & MWI & 182 \\
  Mozambique & MOZ & 121 \\
  Tanzania & TZA & 110 \\
  Burundi & BDI & 109 \\
  Namibia & NAM & 92 \\
  Benin & BEN & 72 \\
  Zambia & ZMB & 71 \\
  Ghana & GHA & 66 \\
  Ethiopia & ETH & 64 \\
  Uganda & UGA & 64 \\
  Cameroon & CMR & 58 \\
  Zimbabwe & ZWE & 58 \\
  Eswatini & SWZ & 51 \\
  South Africa & ZAF & 51 \\
  Senegal & SEN & 45 \\
  Morocco & MAR & 45 \\
  Dem.\ Rep.\ of Congo & COD & 39 \\
  Gabon & GAB & 37 \\
  Niger & NER & 35 \\
  Guinea & GIN & 34 \\
  Mali & MLI & 34 \\
  Angola & AGO & 30 \\
  Burkina Faso & BFA & 30 \\
  Rwanda & RWA & 30 \\
  Chad & TCD & 25 \\
  Central African Republic & CAF & 25 \\
  Madagascar & MDG & 22 \\
  Togo & TGO & 21 \\
  C\^{o}te d'Ivoire & CIV & 19 \\
  Sierra Leone & SLE & 14 \\
  Lesotho & LSO & 11 \\
  Liberia & LBR & 8 \\
  Comoros & COM & 3 \\
  \midrule
  \textbf{Total (35 countries)} & & \textbf{2,166} \\
\end{longtable}

\clearpage
\begin{table}[htbp]
  \centering
  \caption{Last Pre-to-Onset Diagnostic Contrast: TWFE Event-Study DiD}
  \label{tab:pretrend_twfe}
  \begin{tabular}{lrll}
    \toprule
    Sector & $\hat{\beta}_{\ell=-1}$ (ref. $\ell=0$) & 95\% CI & Non-zero contrast \\
    \midrule
  \multicolumn{4}{l}{\textbf{China}}\\
  \quad Agriculture, Forestry and Fishing & 0.306 & [-0.402,~1.014] & No \\
  \quad Communications & 1.660$^{*}$ & [0.701,~2.619] & Yes \\
  \quad Education & 0.283 & [-0.217,~0.783] & No \\
  \quad Emergency Response & -3.000$^{*}$ & [-4.485,~-1.516] & Yes \\
  \quad Energy Generation and Supply & -2.934$^{*}$ & [-3.858,~-2.011] & Yes \\
  \quad Government and Civil Society & -0.286 & [-0.682,~0.110] & No \\
  \quad Health & 1.146$^{*}$ & [0.693,~1.600] & Yes \\
  \quad Other Multisector & 0.883 & [-0.651,~2.418] & No \\
  \quad Other Social Infrastructure and Services & -2.335$^{*}$ & [-3.245,~-1.426] & Yes \\
  \quad Transport and Storage & -0.306 & [-0.751,~0.139] & No \\
  \quad Water Supply and Sanitation & -1.800$^{*}$ & [-2.383,~-1.218] & Yes \\
  \midrule
  \multicolumn{4}{l}{\textbf{World Bank}}\\
  \quad Agriculture, Forestry and Fishing & -0.417$^{*}$ & [-0.632,~-0.202] & Yes \\
  \quad Banking and Financial Services & -0.770$^{*}$ & [-1.217,~-0.323] & Yes \\
  \quad Communications & 2.437$^{*}$ & [1.590,~3.283] & Yes \\
  \quad Education & 0.095 & [-0.228,~0.418] & No \\
  \quad Energy Generation and Supply & -0.164 & [-0.480,~0.152] & No \\
  \quad General Environmental Protection & -0.562$^{*}$ & [-1.065,~-0.058] & Yes \\
  \quad Government and Civil Society & -0.301$^{*}$ & [-0.547,~-0.055] & Yes \\
  \quad Health & 0.243 & [-0.008,~0.494] & No \\
  \quad Industry, Mining, Construction & -0.070 & [-0.306,~0.167] & No \\
  \quad Other Social Infrastructure and Services & -0.160 & [-0.386,~0.066] & No \\
  \quad Transport and Storage & -0.878$^{*}$ & [-1.111,~-0.646] & Yes \\
  \quad Water Supply and Sanitation & -0.475$^{*}$ & [-0.684,~-0.266] & Yes \\
    \bottomrule
  \end{tabular}
  \vspace{4pt}
  \small\textit{Note:} Last pre-to-onset diagnostic contrast at event time
  $\ell=-1$ from the TWFE specification, with $\ell=0$ as the omitted
  reference period.
  $^{*}$ Statistically significant at the 5\% level.
  A non-zero contrast is recorded when the coefficient is significantly
  different from zero, indicating a non-zero last pre-to-onset contrast at $\ell=-1$.
\end{table}

\clearpage
\begin{table}[htbp]
  \centering
  \caption{Last Pre-to-Onset Diagnostic Contrast: dCdH DiD Estimator}
  \label{tab:pretrend_dcdh}
  \begin{tabular}{lrll}
    \toprule
    Sector & $\hat{\beta}_{\ell=-1}$ (ref. $\ell=0$) & 95\% CI & Non-zero contrast \\
    \midrule
  \multicolumn{4}{l}{\textbf{China}}\\
  \quad Agriculture, Forestry and Fishing & -0.159 & [-0.547,~0.229] & No \\
  \quad Communications & -0.020 & [-0.475,~0.435] & No \\
  \quad Education & -0.969$^{*}$ & [-1.243,~-0.695] & Yes \\
  \quad Emergency Response & -1.517$^{*}$ & [-2.066,~-0.968] & Yes \\
  \quad Energy Generation and Supply & -0.020 & [-0.428,~0.387] & No \\
  \quad Government and Civil Society & -0.965$^{*}$ & [-1.230,~-0.700] & Yes \\
  \quad Health & -1.076$^{*}$ & [-1.321,~-0.831] & Yes \\
  \quad Other Multisector & -0.752$^{*}$ & [-1.230,~-0.274] & Yes \\
  \quad Other Social Infrastructure and Services & -2.342$^{*}$ & [-2.736,~-1.948] & Yes \\
  \quad Transport and Storage & -0.208 & [-0.529,~0.113] & No \\
  \quad Water Supply and Sanitation & -0.737$^{*}$ & [-1.027,~-0.447] & Yes \\
  \midrule
  \multicolumn{4}{l}{\textbf{World Bank}}\\
  \quad Agriculture, Forestry and Fishing & -0.220$^{*}$ & [-0.414,~-0.026] & Yes \\
  \quad Banking and Financial Services & -0.331$^{*}$ & [-0.556,~-0.106] & Yes \\
  \quad Communications & 0.332$^{*}$ & [0.009,~0.655] & Yes \\
  \quad Education & -0.684$^{*}$ & [-0.905,~-0.463] & Yes \\
  \quad Energy Generation and Supply & -0.595$^{*}$ & [-0.811,~-0.379] & Yes \\
  \quad General Environmental Protection & -0.486$^{*}$ & [-0.780,~-0.192] & Yes \\
  \quad Government and Civil Society & 0.009 & [-0.174,~0.193] & No \\
  \quad Health & -0.345 & [-0.733,~0.043] & No \\
  \quad Industry, Mining, Construction & 0.087 & [-0.060,~0.234] & No \\
  \quad Other Social Infrastructure and Services & -0.703$^{*}$ & [-0.919,~-0.487] & Yes \\
  \quad Transport and Storage & -0.485$^{*}$ & [-0.645,~-0.325] & Yes \\
  \quad Water Supply and Sanitation & -0.506$^{*}$ & [-0.658,~-0.354] & Yes \\
    \bottomrule
  \end{tabular}
  \vspace{4pt}
  \small\textit{Note:} Last pre-to-onset diagnostic contrast at event time
  $\ell=-1$ from the dCdH specification, with $\ell=0$ as the omitted
  reference period.
  $^{*}$ Statistically significant at the 5\% level.
  A non-zero contrast is recorded when the coefficient is significantly
  different from zero, indicating a non-zero last pre-to-onset contrast at $\ell=-1$.
\end{table}

\clearpage
\small\setlength{\tabcolsep}{2pt}
\begin{longtable}{@{}>{\raggedright\arraybackslash}p{0.24\linewidth}rrr>{\raggedright\arraybackslash}p{0.14\linewidth}>{\raggedright\arraybackslash}p{0.14\linewidth}>{\raggedright\arraybackslash}p{0.14\linewidth}@{}}
\caption{TWFE Event-Study Results}\label{tab:a1} \\
\toprule
\textbf{Sector} & \textbf{$\hat{\beta}_{\ell=+1}$} & \textbf{$\hat{\beta}_{\ell=+2}$} & \textbf{$\hat{\beta}_{\ell=+3}$} & \textbf{95\% CI ($\ell=+1$)} & \textbf{95\% CI ($\ell=+2$)} & \textbf{95\% CI ($\ell=+3$)} \\
\midrule
\endfirsthead
\multicolumn{7}{l}{\tablename\ \thetable{} -- continued} \\
\toprule
\textbf{Sector} & \textbf{$\hat{\beta}_{\ell=+1}$} & \textbf{$\hat{\beta}_{\ell=+2}$} & \textbf{$\hat{\beta}_{\ell=+3}$} & \textbf{95\% CI ($\ell=+1$)} & \textbf{95\% CI ($\ell=+2$)} & \textbf{95\% CI ($\ell=+3$)} \\
\midrule
\endhead
\midrule
\multicolumn{7}{r}{continued on next page} \\
\endfoot
\bottomrule
\multicolumn{7}{p{\linewidth}}{\footnotesize\textit{Notes:} TWFE post-treatment event-study coefficients. Event time $\ell=0$ is the omitted reference period. Outcome: IWI points. Unit of observation: DHS cluster-period. 95\% confidence intervals based on cluster-robust standard errors clustered at the DHS-cluster level.} \\
\endlastfoot
\midrule
\multicolumn{7}{l}{\textit{China}} \\
\midrule
Agriculture, Forestry and Fishing & 0.500 & 0.162 & 0.116 & {[}-0.15, 1.15{]} & {[}-0.96, 1.28{]} & {[}-1.63, 1.86{]} \\
Communications & 1.009 & 0.773 & 0.707 & {[}0.21, 1.81{]} & {[}-0.63, 2.18{]} & {[}-1.24, 2.65{]} \\
Education & -0.283 & -0.216 & 0.008 & {[}-0.79, 0.22{]} & {[}-1.24, 0.81{]} & {[}-1.59, 1.61{]} \\
Emergency Response & 2.474 & 3.476 & 5.358 & {[}1.31, 3.64{]} & {[}1.28, 5.67{]} & {[}1.94, 8.78{]} \\
Energy Generation and Supply & 2.377 & 6.093 & 7.290 & {[}1.47, 3.29{]} & {[}4.27, 7.92{]} & {[}4.65, 9.93{]} \\
Government and Civil Society & 0.658 & 1.079 & 0.742 & {[}0.30, 1.02{]} & {[}0.33, 1.83{]} & {[}-0.32, 1.80{]} \\
Health & -0.631 & -1.863 & -2.265 & {[}-1.03, -0.23{]} & {[}-2.61, -1.11{]} & {[}-3.41, -1.12{]} \\
Other Multisector & -1.512 & -2.341 & -3.455 & {[}-2.53, -0.50{]} & {[}-4.56, -0.13{]} & {[}-6.44, -0.47{]} \\
Other Social Infrastructure and Services & 2.041 & 3.667 & 7.384 & {[}1.26, 2.82{]} & {[}2.21, 5.13{]} & {[}5.12, 9.64{]} \\
Transport and Storage & 0.651 & 1.278 & 1.044 & {[}0.19, 1.12{]} & {[}0.43, 2.13{]} & {[}-0.22, 2.31{]} \\
Water Supply and Sanitation & 0.992 & 2.495 & 5.460 & {[}0.11, 1.87{]} & {[}0.74, 4.25{]} & {[}2.43, 8.49{]} \\
\midrule
\multicolumn{7}{l}{\textit{World Bank}} \\
\midrule
Agriculture, Forestry and Fishing & 0.198 & 0.388 & 0.521 & {[}-0.02, 0.42{]} & {[}0.03, 0.74{]} & {[}-0.01, 1.05{]} \\
Banking and Financial Services & 0.172 & 1.415 & 2.425 & {[}-0.17, 0.52{]} & {[}0.89, 1.94{]} & {[}1.59, 3.26{]} \\
Communications & -2.389 & -4.446 & -7.326 & {[}-2.93, -1.85{]} & {[}-5.66, -3.23{]} & {[}-8.99, -5.66{]} \\
Education & 0.926 & 1.039 & 1.055 & {[}0.65, 1.20{]} & {[}0.59, 1.48{]} & {[}0.46, 1.65{]} \\
Energy Generation and Supply & 0.349 & 0.221 & 0.103 & {[}0.04, 0.66{]} & {[}-0.28, 0.72{]} & {[}-0.64, 0.85{]} \\
General Environmental Protection & 0.499 & 0.591 & -0.549 & {[}0.04, 0.96{]} & {[}-0.20, 1.38{]} & {[}-1.87, 0.77{]} \\
Government and Civil Society & 0.209 & 0.428 & 0.549 & {[}0.01, 0.41{]} & {[}0.11, 0.75{]} & {[}0.14, 0.95{]} \\
Health & 0.682 & 1.379 & 1.683 & {[}0.47, 0.90{]} & {[}1.02, 1.73{]} & {[}1.20, 2.17{]} \\
Industry, Mining, Construction & 0.367 & 0.615 & 0.789 & {[}0.12, 0.61{]} & {[}0.20, 1.03{]} & {[}0.18, 1.40{]} \\
Other Social Infrastructure and Services & 0.625 & 0.952 & 1.151 & {[}0.42, 0.83{]} & {[}0.62, 1.28{]} & {[}0.68, 1.62{]} \\
Transport and Storage & 1.005 & 1.376 & 1.821 & {[}0.79, 1.22{]} & {[}1.04, 1.72{]} & {[}1.35, 2.29{]} \\
Water Supply and Sanitation & 1.050 & 1.663 & 2.286 & {[}0.86, 1.24{]} & {[}1.34, 1.99{]} & {[}1.83, 2.75{]} \\
\end{longtable}

\clearpage
\small\setlength{\tabcolsep}{2pt}
\begin{longtable}{@{}>{\raggedright\arraybackslash}p{0.24\linewidth}rrr>{\raggedright\arraybackslash}p{0.14\linewidth}>{\raggedright\arraybackslash}p{0.14\linewidth}>{\raggedright\arraybackslash}p{0.14\linewidth}@{}}
\caption{dCdH Difference-in-Differences Results}\label{tab:a2} \\
\toprule
\textbf{Sector} & \textbf{$\hat{\beta}_{\ell=+1}$} & \textbf{$\hat{\beta}_{\ell=+2}$} & \textbf{$\hat{\beta}_{\ell=+3}$} & \textbf{95\% CI ($\ell=+1$)} & \textbf{95\% CI ($\ell=+2$)} & \textbf{95\% CI ($\ell=+3$)} \\
\midrule
\endfirsthead
\multicolumn{7}{l}{\tablename\ \thetable{} -- continued} \\
\toprule
\textbf{Sector} & \textbf{$\hat{\beta}_{\ell=+1}$} & \textbf{$\hat{\beta}_{\ell=+2}$} & \textbf{$\hat{\beta}_{\ell=+3}$} & \textbf{95\% CI ($\ell=+1$)} & \textbf{95\% CI ($\ell=+2$)} & \textbf{95\% CI ($\ell=+3$)} \\
\midrule
\endhead
\midrule
\multicolumn{7}{r}{continued on next page} \\
\endfoot
\bottomrule
\multicolumn{7}{p{\linewidth}}{\footnotesize\textit{Notes:} dCdH post-treatment event-study coefficients, reported on the same event-time normalization as TWFE with $\ell=0$ as the omitted reference period. Outcome: IWI points. Unit of observation: DHS cluster-period. 95\% bootstrap confidence intervals (1,000 replications) clustered at the DHS-cluster level.} \\
\endlastfoot
\midrule
\multicolumn{7}{l}{\textit{China}} \\
\midrule
Agriculture, Forestry and Fishing & 1.022 & 0.360 & 1.844 & {[}-0.38, 2.43{]} & {[}-0.14, 0.86{]} & {[}0.68, 3.01{]} \\
Communications & 1.644 & 1.945 & 2.665 & {[}1.36, 1.93{]} & {[}1.55, 2.34{]} & {[}2.14, 3.19{]} \\
Education & 0.440 & 1.515 & 3.302 & {[}0.21, 0.67{]} & {[}1.10, 1.93{]} & {[}2.38, 4.22{]} \\
Emergency Response & 1.995 & 1.641 & 2.375 & {[}1.50, 2.49{]} & {[}0.66, 2.63{]} & {[}0.96, 3.79{]} \\
Energy Generation and Supply & 0.506 & 1.501 & 0.162 & {[}0.19, 0.82{]} & {[}0.91, 2.09{]} & {[}-0.73, 1.05{]} \\
Government and Civil Society & 0.935 & 0.859 & -0.321 & {[}0.53, 1.34{]} & {[}-0.66, 2.38{]} & {[}-3.25, 2.61{]} \\
Health & 0.696 & 0.588 & 1.551 & {[}0.52, 0.87{]} & {[}0.33, 0.84{]} & {[}1.03, 2.07{]} \\
Other Multisector & 0.670 & 0.957 & 2.228 & {[}0.33, 1.01{]} & {[}0.23, 1.68{]} & {[}1.59, 2.87{]} \\
Other Social Infrastructure and Services & 0.836 & 0.957 & 4.442 & {[}0.44, 1.23{]} & {[}0.39, 1.53{]} & {[}3.34, 5.54{]} \\
Transport and Storage & 0.433 & -0.095 & -2.418 & {[}0.14, 0.73{]} & {[}-0.76, 0.57{]} & {[}-3.93, -0.91{]} \\
Water Supply and Sanitation & 0.390 & 0.665 & 7.082 & {[}0.14, 0.64{]} & {[}-0.06, 1.39{]} & {[}0.36, 13.80{]} \\
\midrule
\multicolumn{7}{l}{\textit{World Bank}} \\
\midrule
Agriculture, Forestry and Fishing & 0.268 & 0.331 & -2.179 & {[}0.05, 0.48{]} & {[}-0.82, 1.48{]} & {[}-3.27, -1.09{]} \\
Banking and Financial Services & 0.177 & 0.617 & -0.689 & {[}-0.03, 0.38{]} & {[}-0.07, 1.31{]} & {[}-2.16, 0.78{]} \\
Communications & 0.111 & 0.219 & 0.412 & {[}-0.08, 0.30{]} & {[}-0.12, 0.56{]} & {[}-0.04, 0.86{]} \\
Education & 1.254 & 1.933 & 3.685 & {[}0.96, 1.55{]} & {[}1.14, 2.73{]} & {[}3.11, 4.26{]} \\
Energy Generation and Supply & 0.183 & -0.953 & 0.276 & {[}-0.15, 0.51{]} & {[}-1.68, -0.23{]} & {[}-1.03, 1.58{]} \\
General Environmental Protection & 0.258 & 1.481 & 3.669 & {[}-0.00, 0.52{]} & {[}0.94, 2.02{]} & {[}1.75, 5.59{]} \\
Government and Civil Society & 0.152 & 0.155 & 0.290 & {[}0.04, 0.27{]} & {[}-0.11, 0.42{]} & {[}-0.11, 0.69{]} \\
Health & -0.177 & 0.804 & 4.418 & {[}-0.37, 0.01{]} & {[}-0.42, 2.03{]} & {[}2.26, 6.57{]} \\
Industry, Mining, Construction & 1.023 & 0.445 & 0.642 & {[}0.62, 1.43{]} & {[}0.12, 0.77{]} & {[}-0.20, 1.48{]} \\
Other Social Infrastructure and Services & 0.673 & 2.567 & 3.401 & {[}0.46, 0.89{]} & {[}1.18, 3.95{]} & {[}0.66, 6.14{]} \\
Transport and Storage & 0.386 & 0.514 & 0.776 & {[}0.23, 0.54{]} & {[}0.17, 0.86{]} & {[}0.19, 1.36{]} \\
Water Supply and Sanitation & 0.723 & 0.952 & 1.309 & {[}0.55, 0.89{]} & {[}0.24, 1.67{]} & {[}0.02, 2.59{]} \\
\end{longtable}

\clearpage
\subsection{Covariate-adjusted TWFE panel robustness}

\small\setlength{\tabcolsep}{2pt}
\begin{longtable}{@{}>{\raggedright\arraybackslash}p{0.34\linewidth}>{\raggedright\arraybackslash}p{0.14\linewidth}>{\raggedright\arraybackslash}p{0.14\linewidth}>{\raggedright\arraybackslash}p{0.14\linewidth}>{\raggedright\arraybackslash}p{0.14\linewidth}@{}}
\caption{TWFE Event-Study with Covariate Adjustment}\label{tab:a6} \\
\toprule
\textbf{Sector} & \textbf{$\hat{\beta}_{\ell=-1}$} & \textbf{$\hat{\beta}_{\ell=+1}$} & \textbf{$\hat{\beta}_{\ell=+2}$} & \textbf{$\hat{\beta}_{\ell=+3}$} \\
\midrule
\endfirsthead
\multicolumn{5}{l}{\tablename\ \thetable{} -- continued} \\
\toprule
\textbf{Sector} & \textbf{$\hat{\beta}_{\ell=-1}$} & \textbf{$\hat{\beta}_{\ell=+1}$} & \textbf{$\hat{\beta}_{\ell=+2}$} & \textbf{$\hat{\beta}_{\ell=+3}$} \\
\midrule
\endhead
\endfoot
\bottomrule
\multicolumn{5}{l}{\footnotesize\textit{Notes:} Covariate-adjusted TWFE event-study coefficients. Event time $\ell=0$ is the omitted reference period. Outcome: IWI points. Unit of observation: DHS cluster-period. Entries report point estimates with 95\% confidence intervals. The specification includes six covariates: population density, conflict exposure, natural disasters, election timing, political stability, and leader birthplace.} \\
\endlastfoot
\midrule
\multicolumn{5}{l}{\textit{China}} \\
\midrule
Agriculture, Forestry and Fishing & -0.370 [-0.853,\,0.113] & 0.460 [0.091,\,0.829] & 0.637 [0.151,\,1.124] & 0.570 [-0.095,\,1.236] \\
Communications & 1.107 [0.447,\,1.766] & 1.563 [1.234,\,1.893] & 1.707 [1.306,\,2.108] & 1.897 [1.413,\,2.382] \\
Education & -1.041 [-1.373,\,-0.709] & 0.446 [0.201,\,0.692] & 1.171 [0.793,\,1.549] & 2.092 [1.478,\,2.706] \\
Emergency Response & -1.001 [-1.520,\,-0.482] & 1.988 [1.489,\,2.487] & 1.487 [0.611,\,2.363] & 2.042 [0.788,\,3.297] \\
Energy Generation and Supply & -0.200 [-0.779,\,0.379] & 0.396 [0.068,\,0.724] & 1.354 [0.755,\,1.953] & 0.279 [-0.489,\,1.047] \\
Government and Civil Society & -1.529 [-1.814,\,-1.243] & 1.262 [1.040,\,1.484] & 2.554 [2.148,\,2.961] & 3.112 [2.640,\,3.585] \\
Health & -0.890 [-1.176,\,-0.603] & 0.626 [0.432,\,0.820] & 0.914 [0.647,\,1.180] & 2.288 [1.954,\,2.622] \\
Other Multisector & -1.917 [-2.446,\,-1.389] & 0.599 [0.225,\,0.972] & 1.387 [0.707,\,2.067] & 3.138 [2.544,\,3.732] \\
Other Social Infrastructure and Services & -1.586 [-1.998,\,-1.175] & 0.902 [0.507,\,1.298] & 1.049 [0.402,\,1.696] & 3.258 [2.496,\,4.020] \\
Transport and Storage & -0.754 [-1.083,\,-0.424] & 0.939 [0.673,\,1.206] & 1.975 [1.607,\,2.343] & 2.251 [1.767,\,2.735] \\
Water Supply and Sanitation & -0.776 [-1.129,\,-0.423] & 0.707 [0.450,\,0.963] & 1.386 [0.560,\,2.211] & 3.338 [1.706,\,4.970] \\
\midrule
\multicolumn{5}{l}{\textit{World Bank}} \\
\midrule
Agriculture, Forestry and Fishing & -0.290 [-0.471,\,-0.108] & 0.129 [-0.008,\,0.267] & 0.271 [0.081,\,0.461] & 0.326 [0.065,\,0.587] \\
Banking and Financial Services & -0.184 [-0.468,\,0.100] & -0.028 [-0.284,\,0.227] & 0.840 [0.456,\,1.223] & 1.223 [0.728,\,1.718] \\
Communications & -1.300 [-1.634,\,-0.965] & 0.154 [-0.044,\,0.353] & 0.574 [0.249,\,0.899] & 1.022 [0.572,\,1.472] \\
Education & -0.494 [-0.777,\,-0.212] & 1.125 [0.902,\,1.347] & 1.735 [1.446,\,2.024] & 2.134 [1.805,\,2.462] \\
Energy Generation and Supply & -0.836 [-1.082,\,-0.590] & 0.430 [0.225,\,0.635] & 0.876 [0.610,\,1.141] & 1.000 [0.657,\,1.343] \\
General Environmental Protection & -0.700 [-1.025,\,-0.374] & 0.659 [0.381,\,0.937] & 0.986 [0.555,\,1.417] & 0.162 [-0.453,\,0.776] \\
Government and Civil Society & -0.409 [-0.639,\,-0.179] & 0.420 [0.262,\,0.579] & 0.800 [0.592,\,1.007] & 0.995 [0.753,\,1.237] \\
Health & 0.190 [-0.007,\,0.388] & 0.553 [0.384,\,0.723] & 1.324 [1.071,\,1.577] & 1.494 [1.189,\,1.799] \\
Industry, Mining, Construction & 0.309 [0.124,\,0.494] & 0.191 [0.030,\,0.351] & 0.229 [-0.024,\,0.483] & 0.092 [-0.241,\,0.425] \\
Other Social Infrastructure and Services & -0.390 [-0.580,\,-0.201] & 0.720 [0.560,\,0.880] & 1.284 [1.073,\,1.496] & 1.574 [1.313,\,1.835] \\
Transport and Storage & -0.524 [-0.720,\,-0.329] & 0.860 [0.701,\,1.020] & 0.985 [0.780,\,1.190] & 1.003 [0.755,\,1.251] \\
Water Supply and Sanitation & -0.345 [-0.515,\,-0.175] & 0.856 [0.717,\,0.995] & 1.503 [1.303,\,1.704] & 1.775 [1.517,\,2.033] \\
\end{longtable}

\clearpage
\small\setlength{\tabcolsep}{2pt}
\begin{longtable}{@{}>{\raggedright\arraybackslash}p{0.31\linewidth}>{\raggedright\arraybackslash}p{0.15\linewidth}>{\raggedright\arraybackslash}p{0.24\linewidth}>{\raggedright\arraybackslash}p{0.18\linewidth}@{}}
\caption{dCdH Estimator with Covariate Adjustment}\label{tab:a7} \\
\toprule
\textbf{Sector} & \textbf{Estimate ($\hat{\beta}$)} & \textbf{[95\% Confidence Interval]} & \textbf{Non-zero pre-to-onset contrast} \\
\midrule
\endfirsthead
\multicolumn{4}{l}{\tablename\ \thetable{} -- continued} \\
\toprule
\textbf{Sector} & \textbf{Estimate ($\hat{\beta}$)} & \textbf{[95\% Confidence Interval]} & \textbf{Non-zero pre-to-onset contrast} \\
\midrule
\endhead
\midrule
\multicolumn{4}{r}{continued on next page} \\
\endfoot
\bottomrule
\multicolumn{4}{p{\linewidth}}{\footnotesize\textit{Notes:} Last pre-to-onset diagnostic contrast at event time $\ell=-1$ from the dCdH specification with six covariates where feasible, reported relative to the common omitted reference period $\ell=0$. Outcome: IWI points. Unit of observation: DHS cluster-period. 95\% bootstrap confidence intervals (1,000 replications) clustered at the DHS-cluster level. Entries marked $^{\dagger}$ report the estimator's no-covariate fallback for China Other Social Infrastructure and Services (160), China Agriculture, Forestry and Fishing (310), World Bank Banking and Financial Services (240), and World Bank Industry, Mining, Construction (320). Covariates otherwise match Table~\ref{tab:a6}.} \\
\endlastfoot
\midrule
\multicolumn{4}{l}{\textit{China}} \\
\midrule
Agriculture, Forestry and Fishing$^{\dagger}$ & -0.16 & {[}-0.55, 0.23{]} & No \\
Communications & -0.03 & {[}-0.50, 0.44{]} & No \\
Education & -0.87 & {[}-1.14, -0.59{]} & Yes \\
Emergency Response & -1.23 & {[}-1.78, -0.69{]} & Yes \\
Energy Generation and Supply & -0.15 & {[}-0.56, 0.25{]} & No \\
Government and Civil Society & -1.12 & {[}-1.40, -0.83{]} & Yes \\
Health & -0.94 & {[}-1.18, -0.70{]} & Yes \\
Other Multisector & -0.54 & {[}-1.02, -0.07{]} & Yes \\
Other Social Infrastructure and Services$^{\dagger}$ & -2.34 & {[}-2.74, -1.95{]} & Yes \\
Transport and Storage & -0.23 & {[}-0.55, 0.10{]} & No \\
Water Supply and Sanitation & -0.60 & {[}-0.89, -0.32{]} & Yes \\
\midrule
\multicolumn{4}{l}{\textit{World Bank}} \\
\midrule
Agriculture, Forestry and Fishing & -0.21 & {[}-0.38, -0.03{]} & Yes \\
Banking and Financial Services$^{\dagger}$ & -0.33 & {[}-0.56, -0.10{]} & Yes \\
Communications & 0.33 & {[}0.01, 0.65{]} & Yes \\
Education & -0.70 & {[}-0.92, -0.48{]} & Yes \\
Energy Generation and Supply & -0.48 & {[}-0.70, -0.27{]} & Yes \\
General Environmental Protection & -0.54 & {[}-0.84, -0.25{]} & Yes \\
Government and Civil Society & -0.18 & {[}-0.36, 0.00{]} & No \\
Health & -0.13 & {[}-0.51, 0.24{]} & No \\
Industry, Mining, Construction$^{\dagger}$ & 0.09 & {[}-0.06, 0.23{]} & No \\
Other Social Infrastructure and Services & -0.63 & {[}-0.84, -0.43{]} & Yes \\
Transport and Storage & -0.61 & {[}-0.77, -0.46{]} & Yes \\
Water Supply and Sanitation & -0.52 & {[}-0.67, -0.37{]} & Yes \\
\end{longtable}

\clearpage
\small\setlength{\tabcolsep}{2pt}
\begin{longtable}{@{}>{\raggedright\arraybackslash}p{0.24\linewidth}rrr>{\raggedright\arraybackslash}p{0.14\linewidth}>{\raggedright\arraybackslash}p{0.14\linewidth}>{\raggedright\arraybackslash}p{0.14\linewidth}@{}}
\caption{dCdH Event-Study with Covariate Adjustment}\label{tab:a8} \\
\toprule
\textbf{Sector} & \textbf{$\hat{\beta}_{\ell=+1}$} & \textbf{$\hat{\beta}_{\ell=+2}$} & \textbf{$\hat{\beta}_{\ell=+3}$} & \textbf{95\% CI ($\ell=+1$)} & \textbf{95\% CI ($\ell=+2$)} & \textbf{95\% CI ($\ell=+3$)} \\
\midrule
\endfirsthead
\multicolumn{7}{l}{\tablename\ \thetable{} -- continued} \\
\toprule
\textbf{Sector} & \textbf{$\hat{\beta}_{\ell=+1}$} & \textbf{$\hat{\beta}_{\ell=+2}$} & \textbf{$\hat{\beta}_{\ell=+3}$} & \textbf{95\% CI ($\ell=+1$)} & \textbf{95\% CI ($\ell=+2$)} & \textbf{95\% CI ($\ell=+3$)} \\
\midrule
\endhead
\midrule
\multicolumn{7}{r}{continued on next page} \\
\endfoot
\bottomrule
\multicolumn{7}{p{\linewidth}}{\footnotesize\textit{Notes:} dCdH post-treatment event-study coefficients from the covariate-adjusted specification where feasible, reported on the same event-time normalization as TWFE with $\ell=0$ as the omitted reference period. Outcome: IWI points. Unit of observation: DHS cluster-period. 95\% bootstrap confidence intervals (1,000 replications) clustered at the DHS-cluster level. Entries marked $^{\dagger}$ report the estimator's no-covariate fallback for China Other Social Infrastructure and Services (160), China Agriculture, Forestry and Fishing (310), World Bank Banking and Financial Services (240), and World Bank Industry, Mining, Construction (320).} \\
\endlastfoot
\midrule
\multicolumn{7}{l}{\textit{China}} \\
\midrule
Agriculture, Forestry and Fishing$^{\dagger}$ & 1.022 & 0.360 & 1.844 & {[}-1.09, 3.14{]} & {[}-0.14, 0.86{]} & {[}0.68, 3.01{]} \\
Communications & 1.576 & 1.823 & 2.563 & {[}1.30, 1.86{]} & {[}1.43, 2.22{]} & {[}2.03, 3.10{]} \\
Education & 0.356 & 1.441 & 3.091 & {[}-0.01, 0.72{]} & {[}0.42, 2.46{]} & {[}2.19, 4.00{]} \\
Emergency Response & 1.929 & 1.365 & 2.050 & {[}1.43, 2.42{]} & {[}0.36, 2.37{]} & {[}0.63, 3.47{]} \\
Energy Generation and Supply & 0.353 & 1.016 & -0.459 & {[}0.04, 0.66{]} & {[}0.43, 1.61{]} & {[}-1.38, 0.46{]} \\
Government and Civil Society & 1.427 & 2.470 & 4.964 & {[}1.15, 1.70{]} & {[}1.83, 3.11{]} & {[}-4.72, 14.65{]} \\
Health & 0.663 & 0.563 & 2.188 & {[}0.49, 0.83{]} & {[}0.31, 0.82{]} & {[}1.67, 2.70{]} \\
Other Multisector & 0.682 & 0.994 & 2.262 & {[}0.33, 1.03{]} & {[}0.27, 1.72{]} & {[}1.63, 2.89{]} \\
Other Social Infrastructure and Services$^{\dagger}$ & 0.836 & 0.957 & 4.442 & {[}0.36, 1.32{]} & {[}0.39, 1.52{]} & {[}3.34, 5.54{]} \\
Transport and Storage & 0.552 & 0.044 & -1.197 & {[}0.24, 0.86{]} & {[}-0.65, 0.74{]} & {[}-2.74, 0.34{]} \\
Water Supply and Sanitation & 0.607 & 0.746 & 6.499 & {[}0.38, 0.83{]} & {[}-0.29, 1.78{]} & {[}-0.09, 13.09{]} \\
\midrule
\multicolumn{7}{l}{\textit{World Bank}} \\
\midrule
Agriculture, Forestry and Fishing & 0.440 & 0.727 & -1.497 & {[}0.23, 0.65{]} & {[}-0.59, 2.04{]} & {[}-3.49, 0.49{]} \\
Banking and Financial Services$^{\dagger}$ & 0.177 & 0.617 & -0.689 & {[}-0.03, 0.38{]} & {[}-0.07, 1.31{]} & {[}-2.16, 0.78{]} \\
Communications & 0.168 & 0.310 & 0.392 & {[}-0.01, 0.35{]} & {[}-0.02, 0.64{]} & {[}-0.05, 0.84{]} \\
Education & 1.292 & 1.368 & 3.559 & {[}1.00, 1.58{]} & {[}0.60, 2.13{]} & {[}2.99, 4.13{]} \\
Energy Generation and Supply & 0.005 & -0.934 & 0.530 & {[}-0.32, 0.33{]} & {[}-1.60, -0.27{]} & {[}-0.79, 1.85{]} \\
General Environmental Protection & 0.306 & 1.547 & 3.536 & {[}0.04, 0.57{]} & {[}1.01, 2.09{]} & {[}1.66, 5.41{]} \\
Government and Civil Society & 0.217 & 0.481 & 0.927 & {[}0.10, 0.33{]} & {[}0.21, 0.75{]} & {[}0.53, 1.32{]} \\
Health & -0.138 & 0.750 & 4.654 & {[}-0.35, 0.07{]} & {[}-0.60, 2.10{]} & {[}3.85, 5.46{]} \\
Industry, Mining, Construction$^{\dagger}$ & 1.023 & 0.445 & 0.642 & {[}0.62, 1.43{]} & {[}0.12, 0.78{]} & {[}-0.20, 1.48{]} \\
Other Social Infrastructure and Services & 0.701 & 2.351 & 2.842 & {[}0.49, 0.91{]} & {[}1.00, 3.70{]} & {[}0.05, 5.63{]} \\
Transport and Storage & 0.637 & 0.992 & 1.681 & {[}0.48, 0.79{]} & {[}0.65, 1.33{]} & {[}1.10, 2.26{]} \\
Water Supply and Sanitation & 0.741 & 1.228 & 1.980 & {[}0.57, 0.91{]} & {[}0.54, 1.92{]} & {[}0.70, 3.26{]} \\
\end{longtable}

\clearpage
\subsection{Secondary pooled spatial diagnostics}

\begin{figure}[htbp]
\centering
\includegraphics[width=0.82\linewidth]{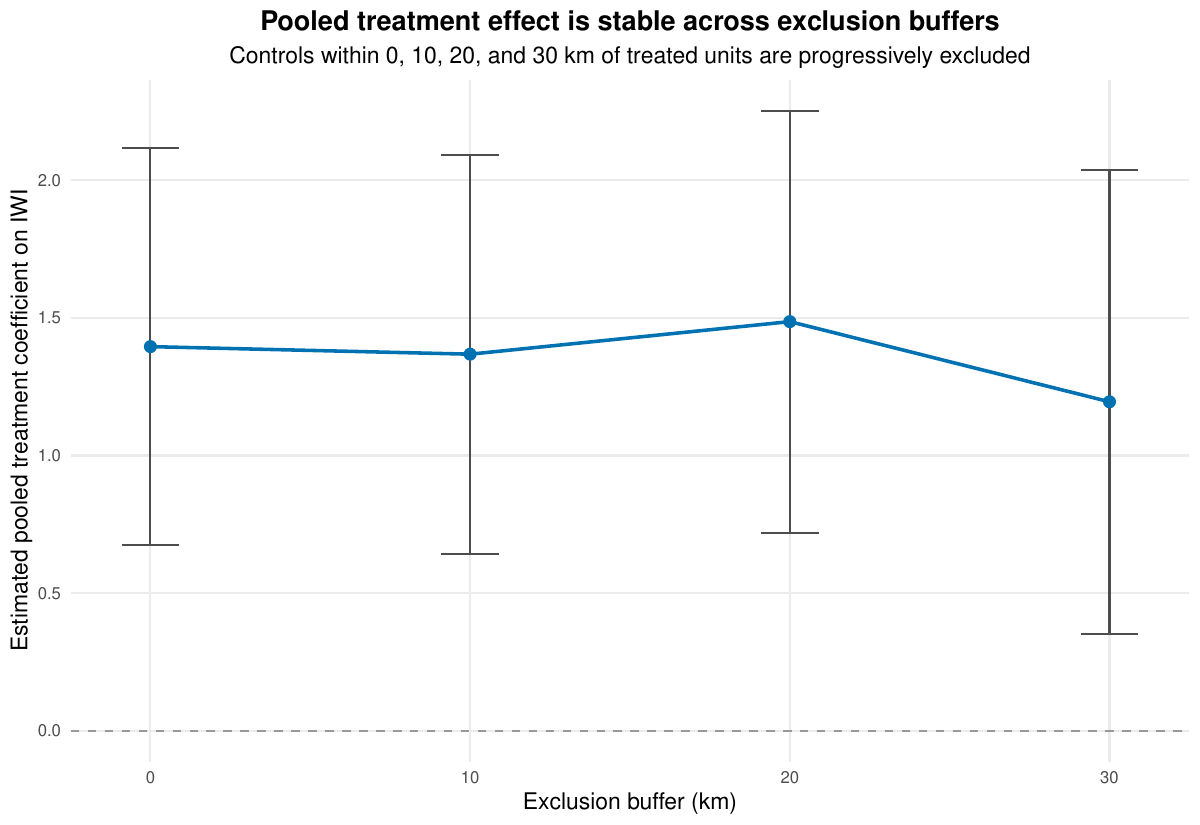}
\caption{Pooled treatment-effect estimates under progressively larger
exclusion buffers around treated units. The close alignment of the
point estimates and confidence intervals is consistent with the conclusion from
Appendix Table~\ref{tab:a4} that nearby-control contamination is not
the primary explanation for the results.}
\label{fig:buffer_robustness}
\end{figure}

\clearpage
\begin{table}[htbp]
  \centering
  \caption{Distance-Based Dose--Response Robustness Test}
  \label{tab:a3}
  \begin{tabular}{lrrrr}
  \toprule
  \textbf{Distance Band (km)} & \textbf{Estimate ($\beta$)} & \textbf{Std. Error} & \textbf{t Statistic} & \textbf{p Value} \\
  \midrule
  0--10 & -2.228 & 1.207 & -1.84 & 0.065 \\
  10--20 & -1.274 & 0.798 & -1.60 & 0.110 \\
  20--50 & -2.661*** & 0.565 & -4.71 & $<$0.001 \\
  \bottomrule
  \end{tabular}
  \begin{flushleft}
  \footnotesize\textit{Notes:} OLS estimates comparing IWI levels across distance bands. Reference: clusters $\geq$50~km from treated areas. * $p<0.05$, ** $p<0.01$, *** $p<0.001$.
  \end{flushleft}
\end{table}

\clearpage
\begin{table}[htbp]
  \centering
  \caption{Exclusion Buffer Robustness Test}
  \label{tab:a4}
  \begin{tabular}{lrrrr}
  \toprule
  \textbf{Exclusion Buffer (km)} & \textbf{Estimate ($\hat{\beta}$)} & \textbf{Std. Error} & \textbf{t Statistic} & \textbf{p Value} \\
  \midrule
  0 & 1.395*** & 0.368 & 3.80 & $<$0.001 \\
  10 & 1.368*** & 0.370 & 3.70 & $<$0.001 \\
  20 & 1.486*** & 0.391 & 3.80 & $<$0.001 \\
  30 & 1.195** & 0.430 & 2.78 & 0.006 \\
  \bottomrule
  \end{tabular}
  \begin{flushleft}
  \footnotesize\textit{Notes:} ATT estimates after excluding control clusters within stated buffer distance (km). * $p<0.05$, ** $p<0.01$, *** $p<0.001$.
  \end{flushleft}
\end{table}

\clearpage
\begin{table}[htbp]
  \centering
  \caption{Global Moran's I Spatial Autocorrelation Diagnostic}
  \label{tab:a5}
  \begin{tabular}{lrrr}
  \toprule
  \textbf{Global Moran's I} & \textbf{Expected Value} & \textbf{Variance} & \textbf{p Value} \\
  \midrule
  0.239 & -0.0005 & 0.000162 & $<$0.001 \\
  \bottomrule
  \end{tabular}
  \begin{flushleft}
  \footnotesize\textit{Notes:} Spatial independence test on OLS residuals. Significant positive Moran's I indicates spatial clustering ($p < 0.001$).
  \end{flushleft}
\end{table}

\clearpage
\section*{Acknowledgements}

\textbf{Mattias Antar} would like to thank Connor Jerzak and Adel Daoud for their supervision, valuable feedback, and guidance throughout this work. Their comments and suggestions substantially improved the manuscript.

The authors also gratefully acknowledges SayedMorteza Malaekeh and members of the AI \& Global Development Lab for their insightful comments, discussions, and constructive suggestions during the development of this project.

\section*{Funding}
This research was supported by the Swedish Research Council.

\section*{Declaration of competing interest}

The authors declare no conflicting or competing interests. 

\section*{Data availability}
Full replication data and code to be made available in a Harvard Dataverse repository. 

\end{document}